\documentclass[a4paper]{article}

\usepackage{a4wide}
\usepackage[UKenglish]{babel}
\usepackage{graphicx,subfigure}
	\graphicspath{{./}{figures/}}
\usepackage[colorlinks=true,linkcolor=blue,citecolor=red]{hyperref}
\usepackage{xcolor}
\usepackage{amssymb,amsthm,empheq,bbold}
\usepackage[inline]{enumitem}
\usepackage[ruled,vlined,linesnumbered]{algorithm2e}
\usepackage{caption} 
	\usepackage{multirow}
	\usepackage{booktabs}

\theoremstyle{remark}

\theoremstyle{definition}

\newcommand{\abs}[1]{\left\lvert#1\right\rvert}
\DeclarePairedDelimiter\ave{\langle}{\rangle}

\allowdisplaybreaks

\def\e{{\varepsilon}}
\def\C{\mathbb{C}}
\def\I{{\rm I}}
\def\N{\mathbf{N}}

\title{\bf{A Statistical Mechanics Approach to Describe Cell Re-orientation under Stretch}}

\author{\Large{N. Loy, L. Preziosi 
}}

\date{\today}

\begin{document}

\maketitle

\begin{center}
\textbf{Abstract}\\
\end{center}

Experiments show that when a monolayer of cells cultured on an elastic substrate is subject to a cyclic stretch, cells tend to re-orient either perpendicularly or at an oblique angle with respect to the main direction of the stretch. Due to stochastic effects, however, the distribution of angles achieved by the cells is broader and, experimentally, histrograms over the interval $[0, 90^\circ]$ are reported.
Here we will determine the evolution and the stationary state of probability density functions describing the statistical distribution of the orientations of the cells using Fokker-Planck equations derived from microscopic rules for the evolution of the orientation of the cell. As a first attempt, we shall use a stochastic differential equation related to a very general elastic energy and we will show that the results of the time integration and of the stationary state of the related forward Fokker-Planck equation compare very well with experimental results obtained by different researchers. 
Then, in order to model more accurately the microscopic process of cell re-orientation, we consider discrete in time random processes that allow to recover Fokker-Planck equations through the well known technique of quasi-invariant limit.  In particular, we shall introduce a non-local rule related to the evaluation of the  state of stress experienced by the cell extending its protrusions, and a model of re-orientation as a result of an optimal control internally activated by the cell. Also in the latter case the results match very well with experiments.



\section{Introduction}

In the 80's the study of cardiovascular diseases led to the need of understanding the behaviour of cells of the heart and of the arterial walls subject to periodic deformations due to pulsatile heart contraction and consequent blood flow \cite{Buck1, Buck2}. 
In order to mimick this environment, many authors seeded cells on a substratum that was stretched periodically (see, for instance, the recent review \cite{review} and references therein). It was generally found that for sufficiently high stretching frequencies 
(see \cite{Greiner_2015,Hsu,Jungbauer,Lee_2010,Tondon1})
and amplitudes 
(see  \cite{Boccafoschi, DartschHammerleBetz, Kaunasnew, Mao, Morita}), 
cells tend to align perpendicularly to the main stretching direction or at oblique and symmetric angles with respect to it. This fact well correlates with the observation that smooth muscle cells in the intima of arterial walls are oriented obliquely with respect to the vascular axial direction forming helical-like structures characterizied by an angle with the longitudinal axis between $20^\circ$ and $40^\circ$  \cite{Rhodin, Shirinsky}.

The re-orientation dynamics in vitro is quite robust with respect to both cell type and experimental set-up. In fact, regarding the former aspect, fibroblasts, muscle-type cells, epithelial cells, endothelial cells, osteoblasts, melanocytes, mesenchymal stem cells, all respond in a similar way when periodically stretched.
Regarding the latter aspect, the final result seems to be nearly independent from the applied frequency and amplitude and from the mechanical characteristics of the substrate, with transitions when the corresponding values are smaller that some thresholds, i.e., too low frequencies, too small deformations, too soft substrata. On the other hand, the strain ratio in the two perpendicular directions turns out to be relevant, as well described by the experiments performed by Livne et al.  \cite{Livne}.

From the viewpoint of mathematical modelling, the first attempts to describe the phenomenon were based on a strain avoidance principle, consisting in the assumption that cells tend to re-orient in the direction of minimal strain \cite{Barron,Faust, Morioka, Wang_theor, Wang1995}. 

Successively, it was hypothesized that rather than minimal strain, the main reorientation direction tends to minimize stress
\cite{De, De_2, Livne}. Then, the evolution of the cell orientation $\theta$ is related  to a linear elastic energy $\mathcal{E}$ through 
\begin{equation}\label{eq:1}
\dfrac{d}{dt}\theta \propto -\dfrac{\partial}{\partial \theta}\mathcal{E}.
\end{equation}
In particular, Livne et al. \cite{Livne} model the ensemble of cells on the substratum as a linear elastic anisotropic material subject to biaxial strain and identified the equilibrium orientations $\theta_{eq}$ formed by the cell major axis or of the stress fibers and the direction of stretching having minimal energy. In this way, they found a linear relationship between $\cos^2\theta_{eq}$ and a parameter quantifying the biaxiality of the deformation and the cell's anisotropic material coefficients. They also showed that in this parameter plane, data obtained using fibroblasts tend to align along a straight line and were able to identify the relative slope through a match with experimental data.

Starting from the observation that the experimental results holded true even for deformation ranges  that make questionable the use of linear elasticity (they can go up to 30\% \cite{Faust, Livne}), Lucci and Preziosi in \cite{Lucci} proved that a generalization of the linear relationship found by Livne et al. \cite{Livne} also holds for a very large class of nonlinear constitutive orthotropic models. In the nonlinear framework, the squared cosine of the orientation angle is linearly dependent on a parameter which is the natural generalization of the one found in \cite{Livne}, with a slope depending on a combination of elastic coefficients characterizing the nonlinear strain energy. A detailed bifurcation analysis is given. Also Lazopoulos and coworkers \cite{Lazopoulos1, Lazopoulos2, Stamenovic} employed a finite elasticity framework to describe stress fibers reorganization in strained cells, although they considered only uniaxial substrate stretching and addressed the problem using a non-convex energy, giving an explanation based on the co-existence of phases.

A viscoelastic model is proposed in \cite{Lucci2} to explain why on the time scale of experiments the reorientation phenomenon does not occur for small frequencies, for instance,  as a consequence of the reorganization of focal adhesions.
A Maxwell-like force-deformation relation was also found by Chen and Gao \cite{Chen_Gao} who focused on the dynamics of single stress fibers and focal adhesions made of catch bonds.

However, it must be noticed that for sake of simplicity most of the models mentioned above work in a deterministic framework, while, as in any biological process, 
randomness characterizes several aspects of the mentioned dynamics, such as the assembly and disassembly of stress fibers and of focal adhesions as well as the activation and response of mechanosensing pathways. 
Some of these aspects are considered in 
\cite{Hsu, Hsu_2010, Kaunas} where the focus is on the stochastic evolution of radially oriented stress fibers around the nucleus when the cell is subject to static and cyclic stretch. 
In \cite{De_static} De focused instead on the stochastic stretch-sensitive bond association and dissociation processes taking also into account the elasticity of the cell-substrate system to predict the orientation and stability of focal adhesions in the presence of static as well as cyclically varying stretches.

From the experimental point of view,
the visible result of such uncertainties reflects in a spread in cell orientation, in the sense that the distribution of the orientations of the cells is not represented by a Dirac delta, but by smoother functions. Actually, the outcome of the experiments is often described using histograms and graphs reporting the distribution of the percentage of cell orientations falling in a partition of angle ranges
(see, for instance, \cite{Barron,Chen3D, Faust, Hayakawa, Hayakawa2, Livne, Mao, Neidlinger, Neidlinger2, Morioka, Wang1995, Wangmicrogrooved}). The degree of spreading is not constant but depends on the amplitude and frequency of imposed stretch. Specifically, it increases when decreasing amplitude and frequency. 

The inclusion of some randomness allows the models in \cite{Barron,Chen3D, Morioka, Wang1995} to compare the histograms obtained from the experiments with the curves obtained by the results of simulations of the orientation model. However, there, an analytical distribution function was not provided and the effect of stochasticity was not explored in detail.



One of the first analitycal treatments of the problem of describing the \textit{probability density function} of the orientations of the cells (its time evolution or, at least, the stationary state) is provided by Kemkemer and coauthors \cite{KemkemerBioPh2006,Kemkemer1999}. They express the evolution of the orientation of a cell by the means of an automatic controller, i.e. an ODE describing the temporal evolution of the orientation with a empirical  forcing term that has the desired symmetry. They gain a stochastic differential equation (SDE) by adding a diffusion, and obtain the evolution of the probability density distribution as a backward equation of the SDE. They can easily compute the stationary state of the resulting Fokker-Planck equation, represented by an  exponential of a doubly-wrapped cosine, that is a Boltzmann-like distribution. In particular, they compare the analytical findings with experimental results and show that the Boltzmann-like distributions can describe cell orientations on curved substrates.

As a consequence, many authors consider a Boltzman probability density function $f$
\[
f(t,\theta) \propto e^{-\dfrac{\mathcal{E}(\theta)}{kT}}
\]
that is coherent with the fact that the cells' orientation evolves according to \eqref{eq:1}.
Then, all the effort lies in the modelling of the energy $\mathcal{E}$ of the system and of its temperature $T$.
For example, starting from their already mentioned works \cite{De, De_2}, Safran et al. \cite{SafranDe} describe the cell as a re-orienting dipole subject to a periodic stretch and model the distribution of the orientations as a Boltzmann-distribution with a competition between the force determining the free energy of the dipole and the effective temperature. Faust et al. \cite{Faust}  use this distribution assuming an ${\mathcal{E}}$ corresponding to the strain avoidance hypothesis. Also Mao et al. \cite{Mao} consider a Boltzmann-like distribution with an energy that is the sum of three contributions given by the work done by focal adhesions, pulling force and the elastic potential energy of bars in the tensegrity structure, that however presents a flaw.  

Here we will determine the evolution and the stationary state of probability density functions describing the statistical distribution of the orientations of the cells using Fokker-Planck equations, starting from microscopic rules. In order to do that,  after recalling in Section 2 the mechanical background proposed by Lucci and Preziosi \cite{Lucci}, as a first step we shall model the evolution of the cell orientation by the means of a stochastic differential equation in which the evolution of the direction is related to a general elastic energy plus a stochastic fluctuation (Section 3). In the same section the evolution of the probability density function is, then, classically obtained by the means of a forward equation, namely a Fokker-Planck equation.
We will find the stationary state and prove that it is an asymptotic equilibrium (Section 3.1). We will then show in Section 3.2  that using the elastic energy proposed in \cite{Lucci} the results of the integration of the Fokker-Planck equation and its stationary state compare very well with the experimental results reported in \cite{Faust, Hayakawa, Jungbauer, Livne, Mao}.

In Section 4, we shall describe the process of re-orientation as a discrete in time stochastic process that happens with a certain frequency. In Section 4.1 we will then exploit classical tools of kinetic theory that allow to recover a Boltzmann kinetic equation describing the evolution of the statistical distribution of the orientations of cells. Furthermore, by means of the well known technique known as \textit{quasi-invariant limit}, in Section 4.2 we shall recover a Fokker-Planck equation. Eventually, in Section 4.3 we specify several microscopic rules that allow to derive different Fokker-Planck equations suited to describe the evolution of a probability density function of the cells' orientation. Specifically, we first intoduce a local evaluation of the elastic energy that leads to the same Fokker-Planck equation used in Section 3. Then, we introduce a non-local evaluation of the elastic energy describing the fact that a cell feels the sorrounding state of stress extending their protrusions. In Section 4.3.2 we then propose a model of re-orientation as a result of an internal optimal control problem activated by the cell. In the latter case we compare the results of the integration of the derived Fokker-Planck equation and of its stationary state, obtaining an even better agreement with respect to Section 3.2.

\section{Mechanical Backgrounds}
We consider a two-dimensional substratum seeded by cells at a sub-confluent density that is stretched biaxially. We define the $x$-axis along the direction subject to the maximum stretch, so that the principal strains are along the $x$- and $y$-axes.

For sake of simplicity, we assume that cells behave elastically with elastic energy $\mathcal{U}$, are much softer than the substratum and strongly adhere to it, so that the strain in the specimen is perfectly transferred to cells and is homogeneous with the Cauchy-Green strain tensor in the plane that writes as $\mathbb{C}=\textrm{diag}(\lambda_x,\lambda_y)$. Elasticity allows to describe the asymptotic response considering a constant mean strain (see, for instance, \cite{De, De_2, Lazopoulos1, Lazopoulos2, Livne,Lucci, Stamenovic}).
Effect of substratum deformability, of adhesion remodelling, and of viscoelasticity in cell behaviour is neglected here and considered in a deterministic fashion in \cite{Lucci2,Tensegrity}.

We will denote by $\theta \in [0, 2\pi)$ the cell orientation angle with respect to the $x$-axis. 
Since a cell does not have a real polarization given by a head and a tail (see, for instance, \cite{Wang1995}), configurations with cells aligned along $\theta$ and $\theta+\pi$ are geometrically indistinguishable and therefore also equivalent from the energetic point of view, i.e. $\mathcal{U}(\theta+\pi)=\mathcal{U}(\theta)$. In addition, also the orientation of the axes is equivalent and, as a conseguence, $\mathcal{U}(\pi-\theta)=\mathcal{U}(\theta)$.
So, in conclusion, $\mathcal{U}(\theta)$ is an even $\pi$-periodic function and we can work under the following symmetry requirements 
\begin{itemize}
\item[$U1$:] \qquad${\mathcal{U}}(\theta)={\mathcal{U}}(2\pi-\theta)={\mathcal{U}}(\pi-\theta)={\mathcal{U}}(\pi+\theta), \quad \forall \theta$.
\end{itemize}

Generally speaking, denoting by $\N=(\cos\theta, \sin\theta)$ the orientation direction and $\N_{\bot}=(-\sin\theta,\cos\theta)$ its orthogonal, it is known \cite{Ogden} that an elastic energy density ${\cal U}$ for an orthotropic material can depend on the invariants
\begin{equation}\label{Invariants}
\begin{array}{ll}
 \I_4   := \N\cdot\C    \N = (\lambda_x-     \lambda_y     )\cos^2\theta+\lambda_y
=\frac{1}{2}\left[(\lambda_x-     \lambda_y     )\cos 2\theta+\lambda_x+\lambda_y\right]\,,  \\[12pt]
 \I_5   := \N\cdot\C^2\N = (\lambda_x^2-\lambda_y^2)\cos^2\theta+\lambda_y^2
=\frac{1}{2}\left[(\lambda^2_x-     \lambda^2_y     )\cos 2\theta+\lambda^2_x+\lambda^2_y\right]\,,  \\[12pt] 
 \I_6   := \N_{\perp}\cdot\C\N_{\perp} =\lambda_x      -(\lambda_x    -\lambda_y    )\cos^2\theta 
=-\,\frac{1}{2}\left[(\lambda_x-     \lambda_y     )\cos 2\theta-\lambda_x-\lambda_y\right]\,,  \\[12pt] 
  \I_7   := \N_{\perp}\cdot\C^2\N_{\perp} =  \lambda_x^2-(\lambda_x^2-\lambda_y^2)\cos^2\theta
=-\,\frac{1}{2}\left[(\lambda^2_x-     \lambda^2_y     )\cos 2\theta-\lambda^2_x-\lambda^2_y\right]\,,  \\[12pt] 
 \I_8   := \N_{\perp}\cdot\C\N  = - (\lambda_x-\lambda_y)\sin\theta\cos\theta
= - \,\frac{1}{2}(\lambda_x-\lambda_y)\sin 2\theta\,,
\end{array}
\end{equation}
in addition to the usual invariants characterizing isotropy  $\I_1:= {\rm tr}\, \C$,  $\I_2:= \dfrac{1}{2}\left[({\rm tr}\, \C)^2-{\rm tr}\, \C^2\right]$, and $\I_3:=\det\C$ that do not depend on the angle.

All invariants but $\I_8$ satisfy a priori the symmetry requirements given in $U1$. So, the conditions dictated by  $U1$ are satisfied if ${\mathcal U}$ is an even function of $\I_8$, e.g.  function of $\I_8^2=\frac{1}{4}(\lambda_x-\lambda_y)^2(1-\cos^2 2\theta)$.

Under this assumption, $\mathcal{U}=U(\cos 2\theta)$ and the equilibrium orientations are trivially identified by
\begin{equation}\label{U'}
U'(\cos 2\theta)\sin 2\theta=0\,.
\end{equation}
Therefore, one always has the trivial equilibria $\theta=0$ and $\theta=\pi/2$ and might have further equilibria for the values of $\theta=\hat \theta$ such that  $U'(\cos 2\hat\theta)=0$.

Stability of these configurations is achieved if they correspond to minima of the elastic energy, and therefore it depends on the positivity of $U''(\cos 2\theta)\sin^2 2\theta - U'(\cos 2\theta) \cos 2\theta$.
So, one has the following general stability conditions: 
\begin{equation}\label{stabcond}
\begin{array}{ccl}
\theta=0 			:\quad &{\rm stable}\quad		 \Longleftrightarrow & U'(1)<0\,,\\[12pt]
\theta=\dfrac{\pi}{2}:\quad &{\rm stable}\quad 		 \Longleftrightarrow & U'(-1)>0\,,\\[12pt]
\theta=\hat\theta :\quad &{\rm stable}\quad			\Longleftrightarrow & U'{}'(\cos 2\hat\theta)>0\,,
\end{array}
\end{equation}
whenever the last position exists.

Though the method that will be used holds for any $\mathcal{U}$, and referring to \cite{Lucci} for a more general discussion, in the simulations we will specify
\[
\mathcal{U}=\dfrac{1}{2}\left[K_{\|}(I_4-1)^2+K_{\bot}(I_6-1)^2+K_s I_8^2\right]\,,
\]
that for the problem of interest can be written in terms of $\theta$ as
\begin{equation}\label{eq:U}
\mathcal{U}=\dfrac{(\lambda_x-\lambda_y)^2}{8}\left[K_{\|}(\cos 2\theta+2\Lambda-1)^2+K_{\bot}(\cos 2\theta-2\Lambda+1)^2+K_s(1-\cos^2 2\theta)\right]
\end{equation}
where $\Lambda=\dfrac{\lambda_x-1}{\lambda_x-\lambda_y}$.

Actually, most of the papers (e.g., \cite{Livne}) work in a linear elastic regime and prefer to define $\lambda_x=1+2\varepsilon$ and $\lambda_y=1-2r\varepsilon$, where $\varepsilon$ is the infinitesimal strain, so that $\Lambda=\frac{1}{1+r}$, or $r=\frac{1-\Lambda}{\Lambda}$. 
Then, in terms of $r$ the elastic energy writes as 
\begin{equation}\label{barU}
\mathcal{U}=K_\|\varepsilon^2 \bar{\mathcal{U}}=\dfrac{K_\| \varepsilon^2}{2}
\left\{\left[(r+1)\cos 2\theta+1-r\right]^2+\tilde{K}_{\bot}\left[(r+1)\cos 2\theta-1+r\right]^2+\tilde{K}_s(r+1)^2
(1-\cos^2  2\theta)\right\},
\end{equation}
where $\tilde{K}_{\bot}=\dfrac{K_{\bot}}{K_{\|}}$ and $\tilde{K}_s=\dfrac{K_s}{K_{\|}}$. 

\begin{figure}[!ht]
\centering
\subfigure[]{\includegraphics[scale=0.8]{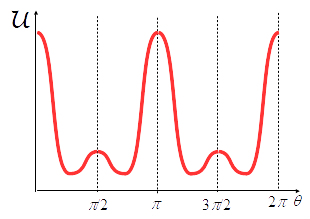}}
\subfigure[]{\includegraphics[scale=0.8]{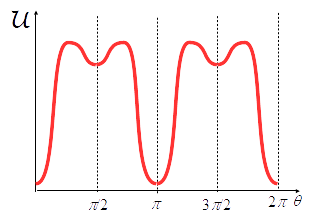}}
\subfigure[]{\includegraphics[scale=0.8]{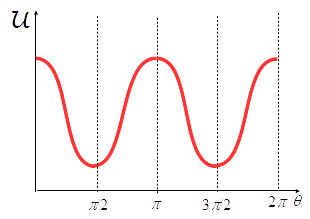}}
\subfigure[]{\includegraphics[scale=0.8]{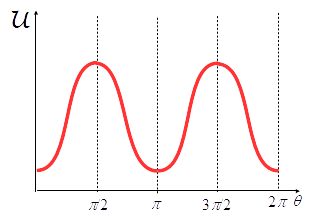}}
\caption{Elastic energy scenarios: (a) corresponds to Case 1, (b) to Case 2, (c) to Case 3 and (d) to Case 4. }
\label{fig:bif}
\end{figure}

Referring to \cite{Lucci,Lucci2} for a more detailed stability analysis, depending on $r$ (or $\Lambda$) and on 
\begin{equation}\label{alpha}
\alpha=\dfrac{1+\tilde{K}_{\bot}-\tilde{K}_s}{1- \tilde{K}_{\bot}}\,,
\end{equation}
defining
$$
\rho(\alpha)=\frac{1+\alpha}{1-\alpha}=\frac{2-\tilde K_s}{\tilde K_s-2\tilde K_\bot},
$$
the following scenarios are possible
\begin{itemize}
\item[{\bf Case 1:}] $\forall r$ if $\alpha>1$ and for $r\in\left[\frac{1}{\rho(\alpha)},\rho(\alpha)\right]$ if $\alpha\in(0,1)$, there is only a stable equilibrium $\theta_{eq}\in\left(0,\frac{\pi}{2}\right)$ such that 
\begin{equation}\label{cos2}
\cos^2\theta_{eq}=\dfrac{1}{2}+\dfrac{1}{\alpha}\left(\dfrac{1}{2}-\dfrac{1}{r+1}\right),
\end{equation}
or 
\begin{equation}\label{cos2theta}
\cos 2\theta_{eq}=\dfrac{1}{\alpha}\,\dfrac{r-1}{r+1}.
\end{equation}
Therefore, due to $U1$, there are four stable equilibria in $[0,2\pi)$, namely in $\theta_{eq}^1=\theta_{eq}$, $\theta_{eq}^2=\pi-\theta_{eq}$, $\theta_{eq}^3=\pi+\theta_{eq}$, $\theta_{eq}^4=2\pi-\theta_{eq}$ (see Fig.\ref{fig:bif}a). 
\item[{\bf Case 2:}] $\forall r$ if $\alpha<-1$ and for $r\in\left[\rho(\alpha),\frac{1}{\rho(\alpha)},\right]$ if $\alpha\in(-1,0)$, there are four stable equilibria in $[0,2\pi)$, namely $\theta_{eq}^1=0$, $\theta_{eq}^2=\pi/2$, $\theta_{eq}^3=\pi$, $\theta_{eq}^4=3\pi/2$ (see Fig.\ref{fig:bif}b);
\item[{\bf Case 3:}] for $r<\rho(\alpha)$  if $\alpha\in(-1,0)$ and  $r<\frac{1}{\rho(\alpha)}$  if $\alpha\in(0,1)$, there are two stable equilibria in $[0,2\pi)$, namely $\theta_{eq}^1=\theta_{eq}^2=\pi/2$, $\theta_{eq}^3=\theta_{eq}^4=3\pi/2$ (see Fig.\ref{fig:bif}c);
\item[{\bf Case 4:}] for $r>\frac{1}{\rho(\alpha)}$  if $\alpha\in(-1,0)$ and  $r>\rho(\alpha)$  if $\alpha\in(0,1)$, there are two stable equilibria in $[0,2\pi)$, namely $\theta_{eq}^1=\theta_{eq}^4=0$, $\theta_{eq}^2=\theta_{eq}^3=\pi$ (see Fig.\ref{fig:bif}d).
\end{itemize}


Working in a deterministic framework, on the basis of Lagrangian mechanics arguments we can relate the evolution in time of the orientation angle $\theta$ with the changes in the virtual work $\mathcal{L}$ done by the stress acting on the cell due to stress fiber alignment. 
Considering an overdamped regime, which corresponds to neglecting inertial effects, we can then write
\begin{equation} \label{lagrangian}
0=-\eta\dfrac{d\theta}{dt}-\dfrac{\partial {\mathcal L}}{\partial\theta}\,,
\end{equation}
where $\eta$ is a viscous-like coefficient measuring cell resistance to realignment. In the elastic case Eq. \eqref{lagrangian} reduces to
\begin{equation}\label{eq:micro}
\eta\dfrac{d\theta}{dt}=-\dfrac{\partial \mathcal{U}}{\partial \theta}(\theta,t)\,,
\end{equation}
or
\begin{equation}\label{eq:micro2}
\dfrac{d\theta}{dt}=-\,\dfrac{\varepsilon^2(t)}{\lambda_\theta}\,\dfrac{\partial\bar{\mathcal{U}}}{\partial \theta}(\theta)\,,
\end{equation}
where $\lambda_\theta=\eta/K_\|$ and we have put in evidence that the strain might be time-dependent. 

Referring to \cite{Lucci2} for a more detailed discussion we here observe that the same equation is obtained for a viscoelastic Maxwell-like model in the limit of high frequencies $\omega$ with respect to the inverse of the viscoelastic relaxation time $\lambda$, i.e., $\lambda\omega\gg1$. On the contrary, in the limit $\lambda\omega\ll 1$ viscous effects dominate and a term $\lambda\omega$ appears at the numerator (related to the appearance of a strain rate, i.e., $\e(t)\dot\e(t)$ instead of $\e^2(t)$), so that the effective $\lambda_\theta$ becomes $\frac{\lambda_\theta}{\lambda\omega}$. Considering that $\lambda$ is of the order of one minute for both stress fiber and focal adhesion remodelling \cite{Chen2013, Pasapera2010}, one has that the transition from low to high frequencies occurs for $\omega$ about $0.01-0.1$ Hertz. 

At variance with the previous deterministic description,  as any biological process, cell re-orientation is strongly affected by their stochastic behaviours. From the experimental point of view, then, this leads to a representation of the orientation state of the ensemble of cells in terms of mean, variance and, whenever possible, frequency histograms, as discussed in the following  (see Figs. \ref{fig_Hayakawa}--\ref{fig_Faust}). 
In parallel, from the theoretical point of view, this leads to the need of determining a probability density function describing the statistical distribution of the orientations. For this reason in the following we will introduce a statistical approach.  
 
\section{Statistical description of the orientations of cells under bi-axial stretch}

\noindent In order to describe analytically the statistical distribution of cell orientation, we introduce the probability density function $f=f(t,\theta)$, $\theta \in [0,\pi)$ such that $f(t,\theta)d\theta$ is the fraction of cells having orientation in $[\theta, \theta+d\theta]$ at time $t$. As discussed before, 
the fact that cells have no identifiable head and tail, implies that if a cell is rotated by $\pi$, it is not possible to perceive a difference in cell orientation. Therefore we shall deal with $\pi$-periodic probability density functions $f$, so that $f(t,\theta)=f(t,\theta+k\pi) \, \forall k \in \mathbb{Z}$.
In addition, as a probability density function, $f$ must satisfy  
\begin{itemize}
\item[$F1$:] \qquad $f \ge 0$;
\item[$F2$:] \qquad  $\displaystyle{\int_{0}^{\pi}} f(t,\theta) \, d\theta =1$.
\end{itemize}
Moreover, due to the symmetry related to the choice of the direction of the axes along the principal strain directions, the following property also holds
\begin{itemize}
\item[$F3$:] \qquad  $f(t,0)=f(t,\pi), \, \forall t\ge0$;
\item[$F4$:]  \qquad  $f$ satisfies the same symmetry property as $U1$, i.e. $f(t,\theta)\ :\quad f(t,\pi-\theta)=f(t,\theta)$;
\end{itemize}
 where $F3$ is also implied by the periodic character of the distribution function.



With the aim of taking randomness into account, we may add a stochastic fluctuation to 
\eqref{eq:micro}, 
\begin{equation}\label{eq:ITO_1}
\dfrac{d\theta}{dt}=-\dfrac{1}{\eta}\dfrac{\partial {{\mathcal{U}}}}{\partial \theta} +
\sqrt{\dfrac{\sigma^2}{\lambda_\theta}}\xi
\end{equation} 
where $\xi$ is a Gaussian random variable with zero mean and unitary variance and $\sigma$ takes into account the stochastic fluctuations linked to uncertainties.
The latter may then be more properly rewritten as an Ito process 
\begin{equation}\label{eq:ITO}
d\theta=-\dfrac{1}{\eta}\dfrac{\partial {{\mathcal{U}}}}{\partial \theta}dt +\sqrt{\dfrac{\sigma^2}{\lambda_\theta}}dW_t
\end{equation} 
where $dW_t=\sqrt{t}\xi$ being $W_t$ a  Wiener process.

The Fokker-Planck equation describing the forward evolution of the probability density distribution $f$ of the orientation of the cells that follows the dynamics \eqref{eq:ITO} is then \cite{Risken}
\begin{equation}\label{FP0}
\dfrac{\partial}{\partial t}f(t,\theta)=\dfrac{\e^2(t)}{\lambda_\theta}\,\dfrac{\partial}{\partial \theta}
\left( \dfrac{\partial \bar{\mathcal{U}}}{\partial \theta}(\theta)f(t,\theta)\right)+
\dfrac{1}{2\lambda_\theta}\dfrac{\partial^2 }{\partial \theta^2}\left({\sigma^2} f(\theta,t)\right).
\end{equation}

We observe that though in most experiment $\e(t)=\e(1-\cos\omega t )$, since we are interested in modelling the process of cell re-orientation, as it is classically done in previous discussed elastic models, we will consider the mean strain $\e$ over an oscillation period.

If we nondimensionalize  time by $\bar{t}=\dfrac{t \e^2}{\lambda_\theta}$, then the  Fokker-Planck equation describing the evolution of $\bar{f}(\bar{t},\theta)=f(\bar{t}\lambda_{\theta}/\e^2,\theta)$ reads 
\begin{equation}\label{eq:FP_Ito}
\dfrac{\partial}{\partial \bar{t}}\bar{f}(\bar{t},\theta)=
\dfrac{\partial}{\partial \theta}\left( \dfrac{\partial \bar{\mathcal{U}}}{\partial \theta}\bar{f}(\bar{t},\theta)\right)+
\dfrac{\partial^2 }{\partial \theta^2}\left(\bar \sigma^2 \bar{f}(\bar{t},\theta)\right)
\end{equation}
where $\bar\sigma^2=\dfrac{\sigma^2}{2\e^2}$.

This already puts in evidence that increasing the stretch amplitude decreases the dimensionless diffusion coefficient $\bar\sigma$ leading to a more focused response and more peaked distribution functions, and vice versa.

As already recalled, the inclusion of viscoelastic effects leads to the same results in the high frequency regime. On the other hand, in the low frequency regime, the dimensional analysis modifies because $\e^2$ is formally replaced by 
$\e^2\lambda\omega$. 
So, the effective dimensionless diffusion coefficient is
$\bar\sigma^2=\dfrac{\sigma^2}{2\lambda \omega\e^2}$,
showing that when the imposed frequency decreases it increases leading to broader distribution functions.

\subsection{Trend to equilibrium}
Dropping the $\bar{ } $ over $f$ and $t$ here and henceforth, if we denote by
\begin{equation}\label{flux}
\mathcal{F}[\theta,f(t,\theta)]= \dfrac{\partial \bar{\mathcal{U}}}{\partial \theta}(\theta) f(t,\theta)
+\dfrac{\partial }{\partial \theta}\left(\bar \sigma^2 f(t,\theta)\right)
\end{equation}
then the $\pi$-periodicity of $\bar{\mathcal{U}}$ and  $f$, implies that 
\begin{equation}\label{cb_1}
\mathcal{F}[\pi,f(t,\pi)]=\mathcal{F}[0,f(t,0)]
\end{equation}
In particular, thanks to the differentiability of $\bar{\mathcal{U}}$, the stationary solution 
$f^{\infty}$ of \eqref{eq:FP_Ito}, coupled with an initial condition $f_0$, satisfying $F1,F2, F3$  is found by imposing 
\begin{equation}\label{eq:flux_van}
\mathcal{F}[\theta,f^{\infty}(\theta)]=0,
\end{equation}
where the r.h.s. side is zero because of the boundary conditions \eqref{cb_1}.
Thus, the stationary state of \eqref{eq:FP_Ito}  is 
\begin{equation}\label{eq:stat_state}
f^{\infty}(\theta)=C\exp\left(-\dfrac{\bar{\mathcal{U}}(\theta)}{\bar\sigma^2} \right)
\end{equation}
where $C$ is a normalization constant. We observe that the maxima (resp. minima) of $f^{\infty}(\theta)$ correspond to minima (resp. maxima) of $\bar {\mathcal{U}}$. In particular, recalling that $f$ is defined in $[0,\pi)$, in Cases 3 and 4 there is only a maximum respectively in $\frac{\pi}{2}$ and $0$. So,  in the former case, due to symmetry the mean corresponds to the mode. A similar property can be obtained in the latter case working in the more convenient periodicity interval $\left(-\,\frac{\pi}{2},\frac{\pi}{2}\right]$, otherwise the mean is trivially and misleadingly equal to $\frac{\pi}{2}$.

On the other hand, in Cases 1 and 2, $f^{\infty}(\theta)$ is a bi-modal distribution in $[0,\pi)$ with modes $\theta_{eq}^1,\pi-\theta_{eq}^1$ and $0$, $\frac{\pi}{2}$, respectively.
Actually, for the already mentioned symmetry reasons, usually, the range of angles used to report experimental data is the first quadrant $[0,\pi/2)$ rather than $[0,\pi)$ or  $[0,2\pi)$.  
In this case, then the notion of mean looses its informative role, especially with respect to the mode that, restricted to $[0,\pi/2)$ is $\theta_{eq}^1$ in Case 1.

\bigskip\noindent
\textbf{Remark} We observe that if $\sigma=0$, i.e. there is no stochastic fluctuation in \eqref{eq:ITO}, then the stationary state given by imposing \eqref{eq:flux_van} is a Dirac delta or a weighted sum of Dirac deltas centered in the stable equilibria.
\bigskip

As usually done for the standard Fokker-Planck equation \cite{furioli2017M3AS}, convergence to the stationary state is studied by analyzing the
monotonicity in time of various Lyapunov functionals of the solution. The typical one is the
relative Shannon entropy, that is defined as follows. Let $f,g : I \subset \mathbb{R}\rightarrow \mathbb{R}_+$ denote two probability densities. Then,
the relative Shannon entropy of $f$ and $f^{\infty}$ is defined by the formula
\begin{equation}\label{def:Shannon}
H(f,f^{\infty})=\int_{0}^{\pi}f(\theta,t) \log\left( \dfrac{f(t,\theta)}{f^{\infty}(t,\theta)}\right) \, d\theta\,.
\end{equation}

As periodic boundary conditions \eqref{cb_1} hold, it is straightforward to prove (see \cite{furioli2017M3AS}) that the Shannon entropy monotonically decreases in time towards the stationary state, i.e.
\[
\dfrac{d}{dt}H(f,f^{\infty}) \le 0\quad \textrm{and} \quad \dfrac{d}{dt}H(f,f^{\infty}) = 0 \quad \textrm{iff} \quad f=f^{\infty}.
\]
Therefore, $f^{\infty}$ is an asymptotic global equilibrium state.

\subsection{Statistical description and comparison with experiments}

Usually, dealing with angles requires circular statistics and the definition of trigonometric moments \cite{Mardia}, e.g. the circular mean
\[
\langle\theta(t)\rangle:=\arctan \dfrac{\beta(t)}{\alpha(t)}, \quad \alpha= \int_0^{\pi}\cos \theta f(t,\theta) \, d\theta, \quad \beta= \int_0^{\pi}\sin \theta f(t,\theta) \, d\theta.
\]
However, the symmetry properties of $f$ would always lead to $\alpha=0$ and therefore $\langle\theta(t)\rangle=\frac{\pi}{2}$. For this reason, we will use the following definition restricted to the first quadrant 
\begin{equation}\label{eq:ave.c}
\bar{\theta}_c(t):=\arctan \dfrac
{\displaystyle\int_0^{\pi/2}\sin \theta f(t,\theta) \, d\theta}
{\displaystyle\int_0^{\pi/2}\cos \theta f(t,\theta) \, d\theta},
\end{equation}
even because it better correlates with the definition of average
\begin{equation}\label{eq:ave}
\bar{\theta}_\ell(t):=2\displaystyle\int_0^{\pi/2} \theta f(t,\theta) \, d\theta,
\end{equation}
used in most experimental papers, where the $2$ accounts for renormalization over $[0,\pi/2)$. We will  also use the coherent definition of variance
\begin{equation}\label{eq:var}
\bar{v}_\ell(t):=2\displaystyle\int_0^{\pi/2} (\theta-\bar{\theta}_\ell)^2 f(t,\theta) \, d\theta.
\end{equation}
An index $\infty$ will identify the quantities above computed for the equilibrium distribution $f^{\infty}$.

\begin{figure}[!htbp]
\centering
\includegraphics[scale=0.25]{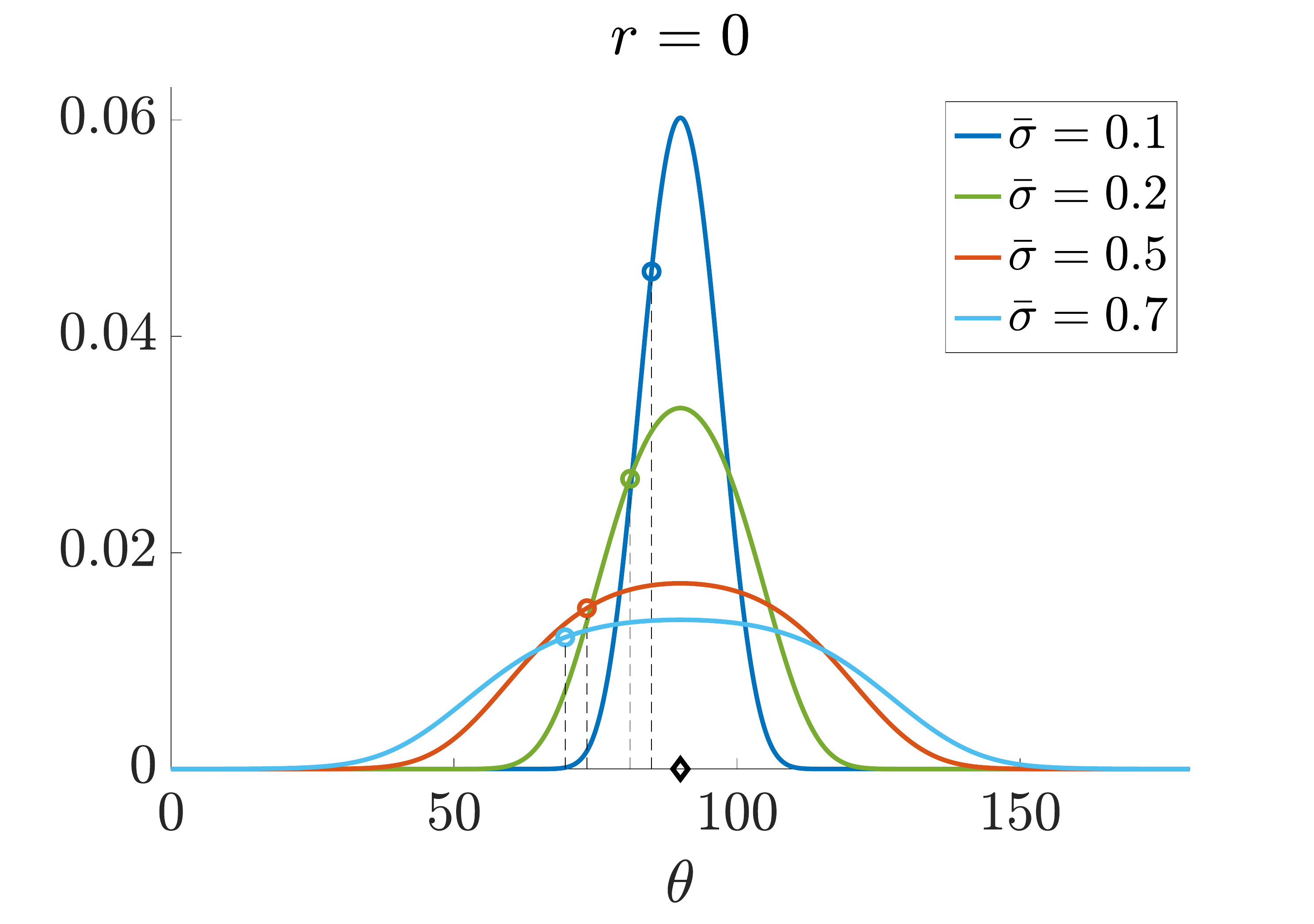}
\hspace{-0.5cm}
\includegraphics[scale=0.25]{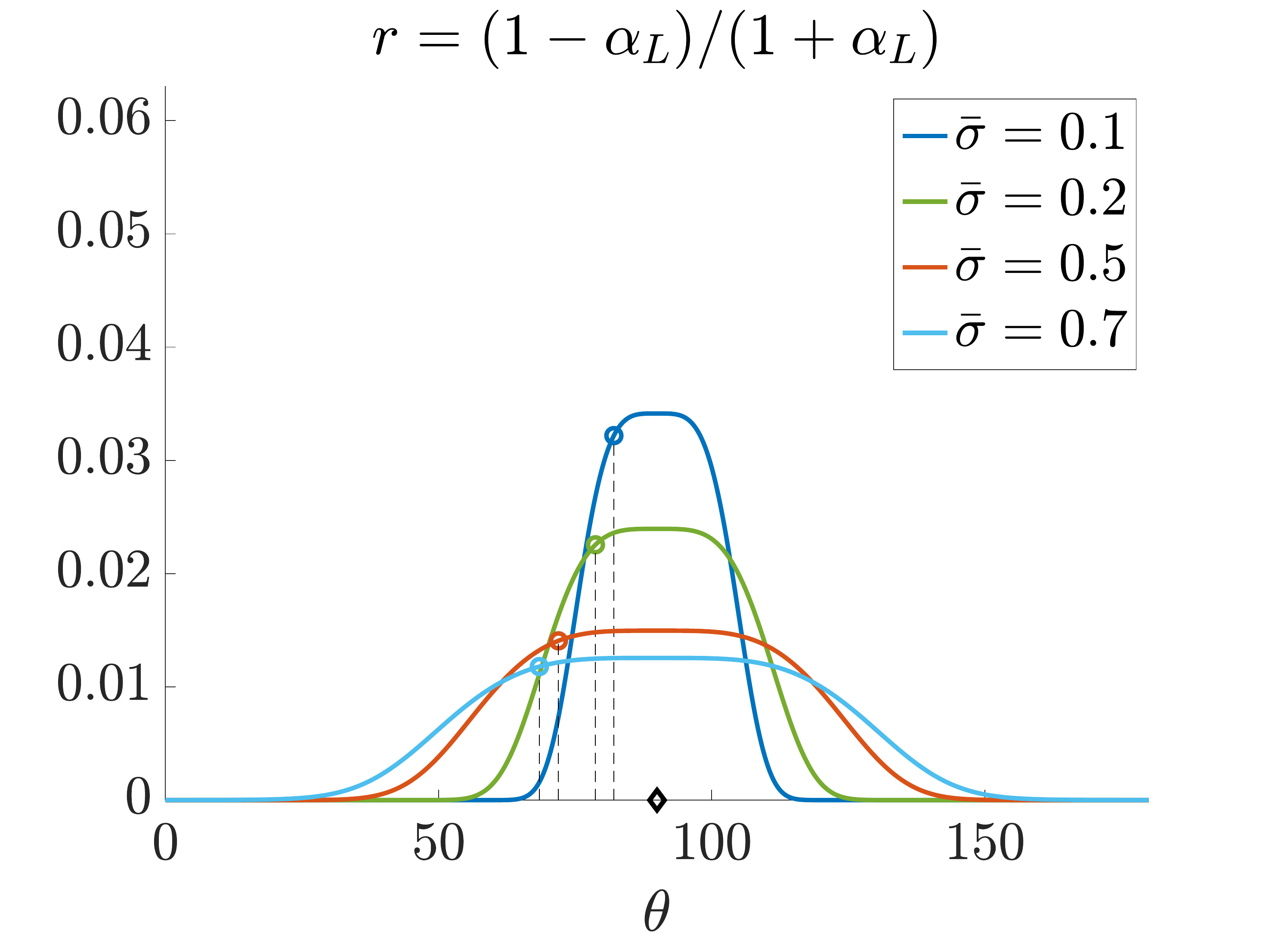}
\\
\includegraphics[scale=0.25]{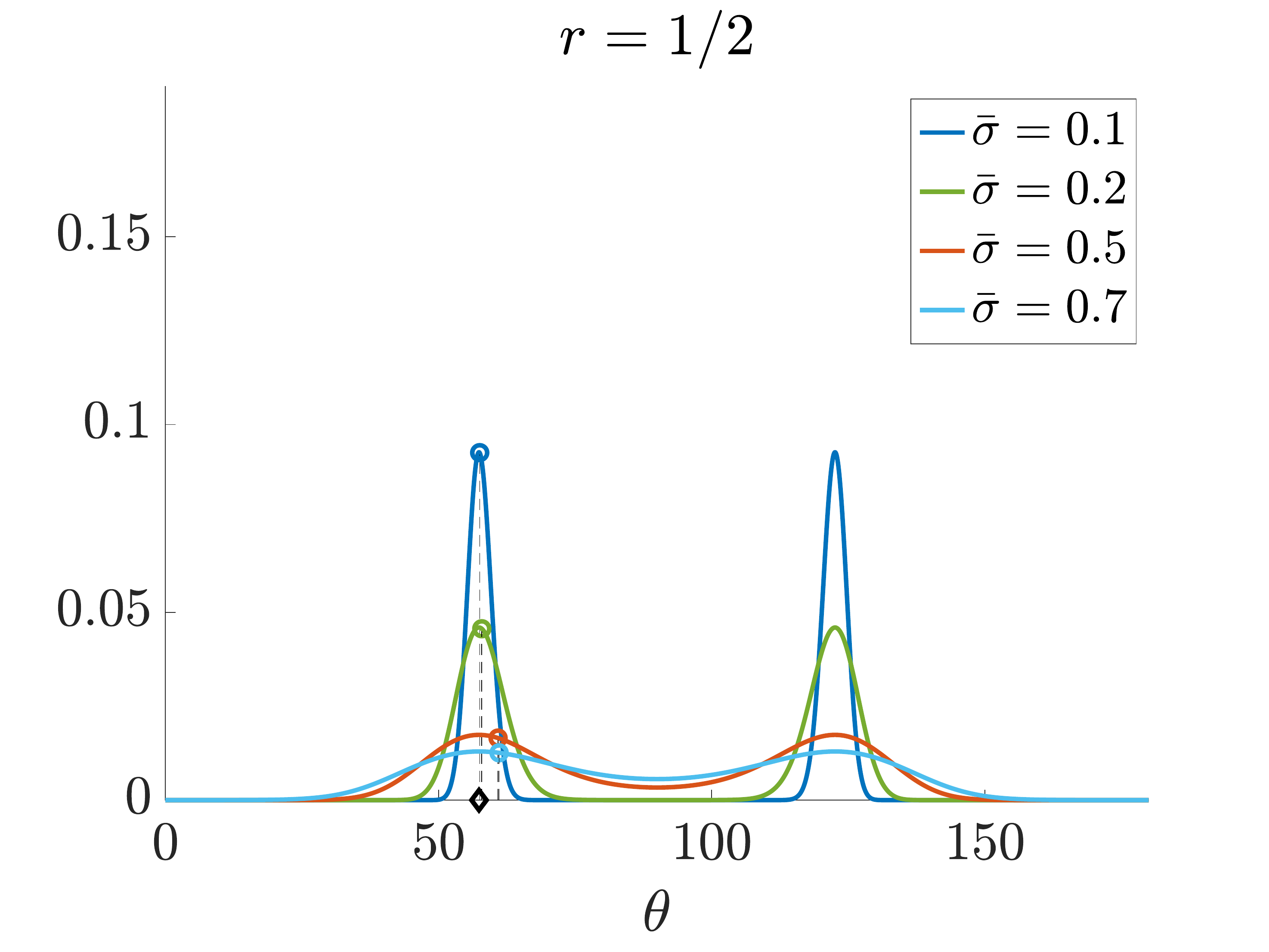}
\hspace{-0.5cm}
\includegraphics[scale=0.25]{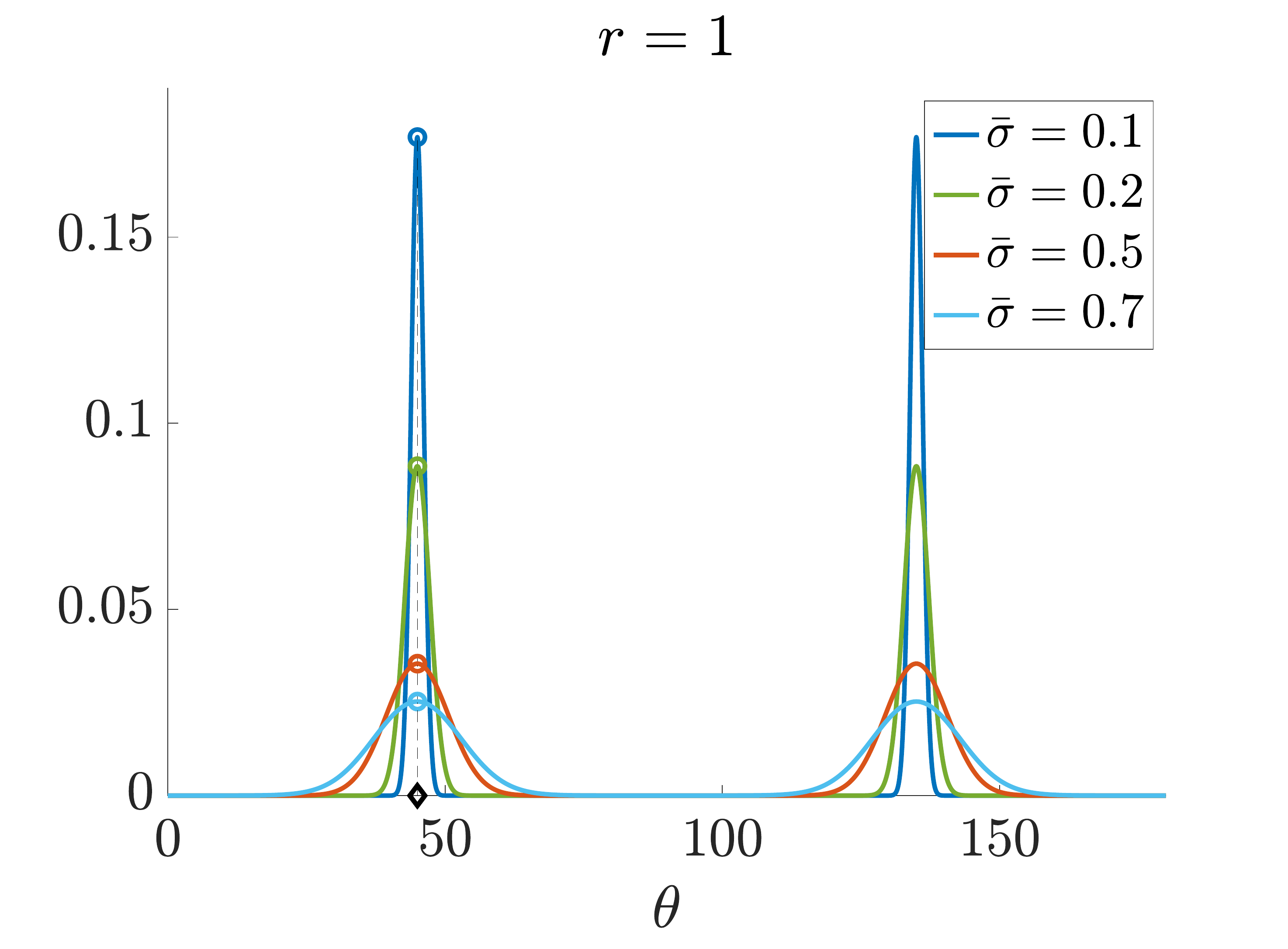}
\\
\begin{tabular}{llc|c|c|c}
\toprule
\multirow{2}*{$r$} & \multirow{2}*{$\theta_{eq}^1$} & \multicolumn{4}{c}{$\bar{\sigma}$} \\
\cmidrule(lr){3-6}
 &  & 0.1 & 0.2 & 0.5 & 0.7 \\
 \cmidrule(lr){3-6}
   & & \multicolumn{4}{c}{$\bar{\theta}_{\ell}^{\infty}$} \\
   \cmidrule(lr){3-6}
$0$ & $90^\circ$ & $84.9^\circ$ & $81.1^\circ$ & $73.4^\circ$ & $69.6^\circ$ \\
$0.115$ & $90^\circ$ & $82.1^\circ$ & $78.7^\circ$ & $71.9^\circ$ &$68.4^\circ$ \\
$0.5$& $57.4^\circ$ & $57.5^\circ$ & $57.9^\circ$& $60.9^\circ$ & $68.4$  \\
$1$ & $45^\circ$ & $45^\circ$& $45^\circ$& $45^\circ$& $45^\circ$ \\
\bottomrule
\end{tabular}
\caption{Profile of the stationary state \eqref{eq:stat_state} with $\bar{\mathcal{U}}$ given by \eqref{barU} for various values of $\bar\sigma$ and $r$ as specified in the title and legend of the figures. In all figures $\tilde{K}_{s}=0.7$. The value $r=\rho(\alpha_L)=\frac{1-\alpha_L}{1+\alpha_L}=0.115$ refers to the bifurcation point.  The table reports $\theta_{eq}^1$ (denoted by a $\boldsymbol{\diamond}$ in the figures) obtained by \eqref{cos2} and the mean $\bar{\theta}_{\ell}^{\infty}$ over $\left[0,\frac{\pi}{2}\right)$ (denoted by a circle in the figures), computed using \eqref{eq:ave} with $f^{\infty}$ defined by \eqref{eq:stat_state}. 
}
\label{fig1}
\end{figure}

\begin{figure}[!htbp]
\centering
\includegraphics[scale=0.3]{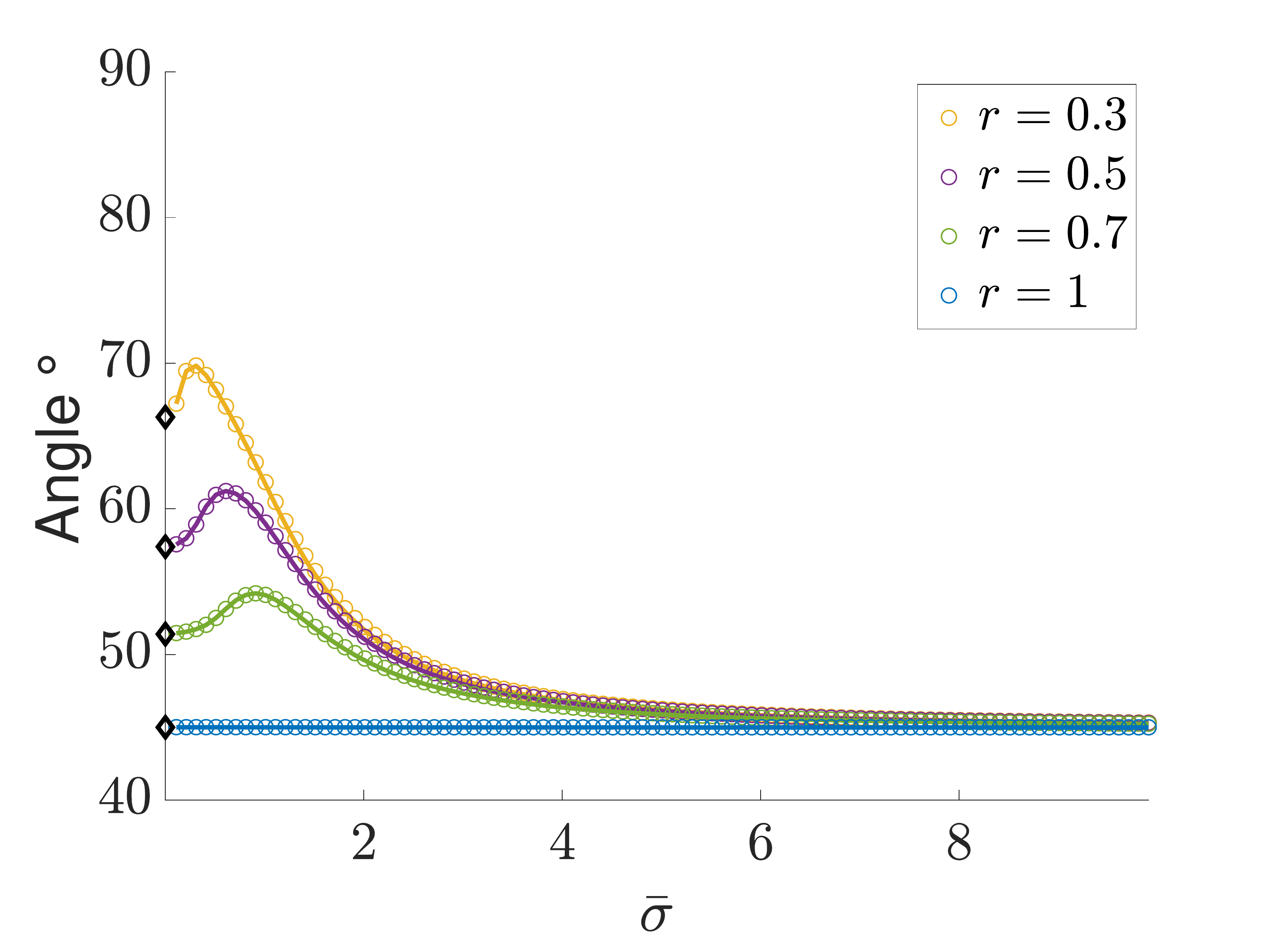}
\caption{Average orientation as a function of $\bar\sigma$ and for different values of $r$. $\bar{\theta}_c^{\infty}$ (circles) and $\bar{\theta_\ell}^{\infty}$ (full line) are computed, respectively, using \eqref{eq:ave.c}-\eqref{eq:ave} and \eqref{eq:stat_state}. Linear and circular average coincide. Moreover, increasing values of $\bar\sigma$ lead to $\pi/4$, corresponding to uniform distributions, while for small values of $\bar\sigma$ the average tends to $\theta_{eq}^1$ (marked by $\boldsymbol{\diamond}$).}
\label{fig_medie}
\end{figure}

However, some remarks are needed. First of all, we observe that in general the average and the mode do not coincide, i.e. $\bar{\theta}^{\infty}_c, \bar{\theta}^{\infty}_\ell\neq \theta_{eq}^1$. They obviously do  when $\sigma\to 0$. 
However, we will see numerically (see Fig. \ref{fig_medie}) that in most cases  $\bar{\theta}_c^{\infty}=\bar{\theta}^{\infty}_\ell$.
In order to clarify this point, in Fig. \ref{fig1} we plot the equilibrium distribution \eqref{eq:stat_state} over the interval $[0,\pi)$ for different values of the parameters $r$ and $\tilde{K}_s$, being $\alpha$ fixed to the value $\alpha_L=0.794$ determined fitting the data of the experiments in Livne et al. \cite{Livne}. 
Then we vary $\tilde{K}_s$ and, from \eqref{alpha}, set
\begin{equation}
\tilde{K}_{\bot}=\dfrac{\tilde{K}_s-1+\alpha_L}{1+\alpha_L}.
\end{equation}
The positivity of $\tilde{K}_{\bot}$ prescribes the compatibility condition 
$$
\tilde{K}_s>1-\alpha_L.
$$

We remark that in Fig. \ref{fig1} and in all the others we preferred to describe angles in degrees rather than in radians for a better readability and a more direct comparison with the statistical descriptions of the experimental results.

In addition to the obvious observation that the diffusion parameter $\bar\sigma$ influences the spread of the orientations, other 
two facts linked to the presence of the diffusion stochastic term emerge explicitly and are put in evidence in Fig. \ref{fig_medie}:
\begin{itemize}
\item unless for the symmetric case $\theta_{eq}^1=\frac{\pi}{4}$ that is always obtained for $r=1$ (see Eq.\eqref{cos2}), the average of the probability density distribution computed over $\left[0,\frac{\pi}{2}\right)$ does not correspond to $\theta_{eq}^1$, that is identified by the mode  in the first quadrant, i.e. the maximum of the distribution function;
\item the average of the probability density depends on $\sigma$ and tends to the mode $\theta_{eq}^1$ (marked by ${\bf \diamond}$ when $\sigma\to 0$ and to $\pi/4$, corresponding to a uniform distribution, when $\sigma\to +\infty$.
\end{itemize}
 In Fig. \ref {fig_medie} we also observe that the linear and the circular average at the stationary state coincide. Therefore, as experiments always consider the linear average, then in the following we shall make reference to $\bar{\theta}_\ell$.

It is evident that in Case 3 when $\theta^1_{eq}=\theta^2_{eq}=\frac{\pi}{2}$, then it is more proper to use $\langle\theta\rangle$, rather than $\bar\theta_l$.

\begin{figure}
\centering
\includegraphics[width=0.45\textwidth]{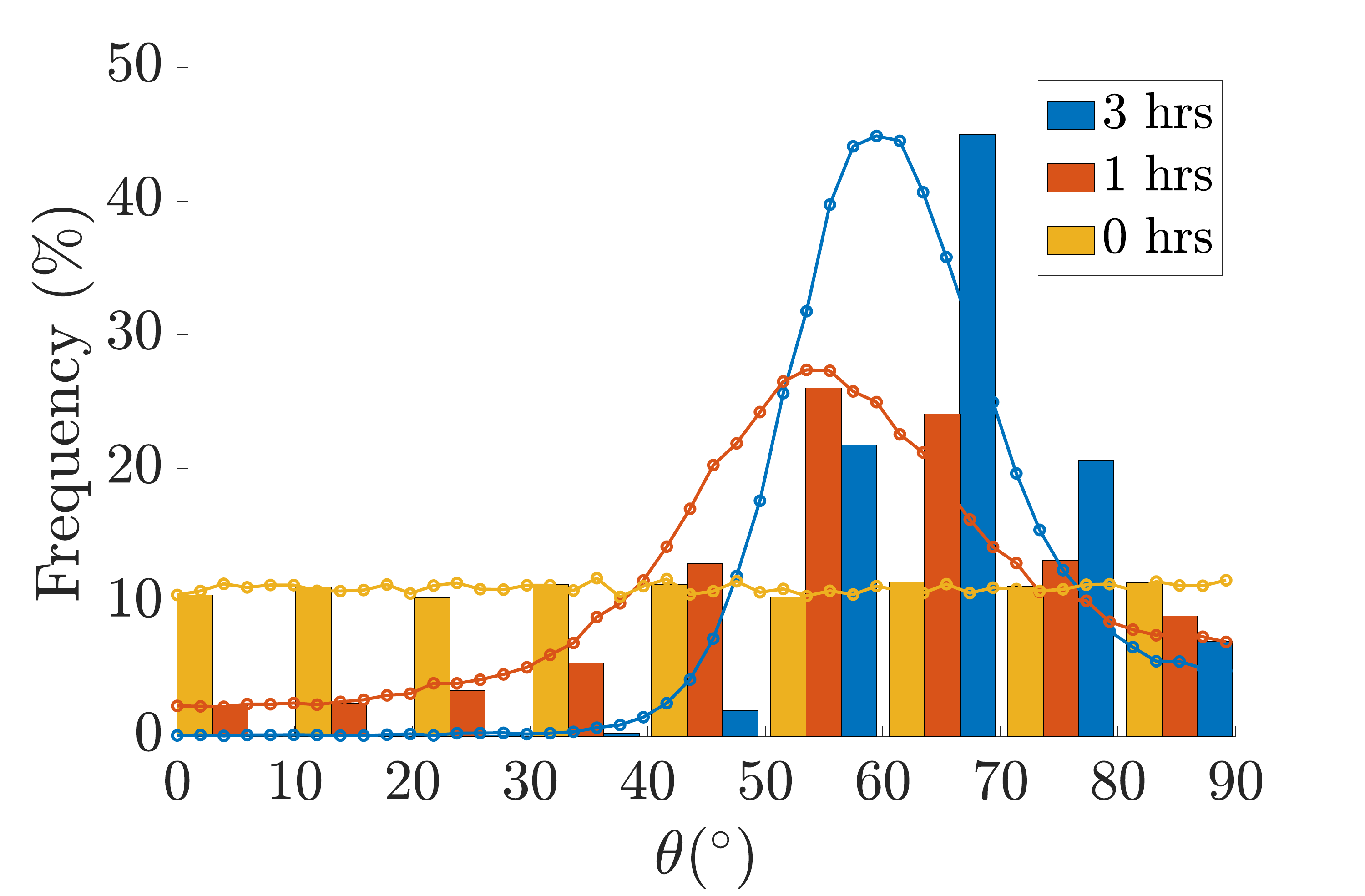}
\includegraphics[width=0.45\textwidth]{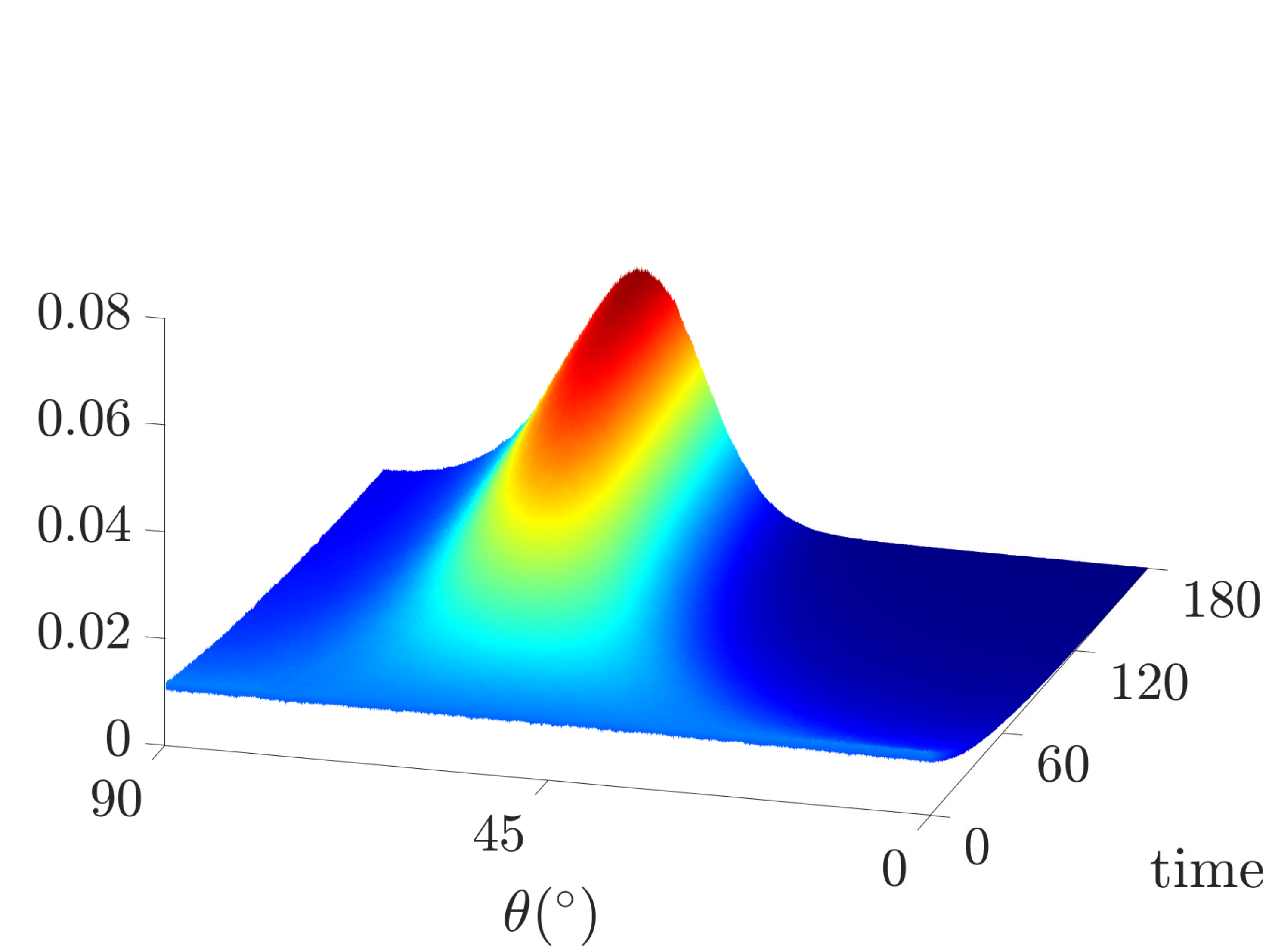}
\caption{Comparison of the evolution of the probability density function obtained by performing a Montecarlo simulation of \eqref{eq:FP_Ito} with the experimental data reported in \cite{Hayakawa}. 
In particular, $\varepsilon=20\%$ and $r=0.4$. Bars refer to experimental data at $t=0,1,3$ hours, respectively in yellow, red and blu. Curves are the recovered probability density functions at $t=0,1,3$ hours. Solution for $\theta_{eq}^1\approx 61^\circ$, $\sigma\approx 0.04$, and $\lambda_\theta\approx 0.18$\,s. On the right, evolution of the probability density function.}
\label{fig_Hayakawa}
\end{figure}

With the aim of comparing the probability density functions with experimental results, we now focus on some papers reporting histograms of the percentage of cells in intervals of orientation angles. As in most cases esperimental data are given for $\theta\in\left[0,\frac{\pi}{2}\right]$, we will restrict to the first quadrant. 

In Fig. \ref{fig_Hayakawa} we compare the temporal evolution of the probability density distribution obtained by integrating \eqref{eq:FP_Ito} with the experimental data reported in \cite{Hayakawa} for $\varepsilon=20\%$, $r=0.4$ and $\omega=1$\,Hz, that implies that we are in a high frequency regime. 
In these experiments it is found that at $t=1$\,h the average orientation is $52.8^{\circ}$, while at $t=3$ hours, when more than the $80\%$ of the cells are oriented at angles of $50^{\circ}$-$80^{\circ}$, the average orientation is $62.02^{\circ}$.  Using \eqref{cos2} and $\alpha=\alpha_L$ the minimum of the elastic energy is obtained at $\theta_{eq}^1\approx 61^\circ$. In particular, in order to integrate \eqref{eq:FP_Ito} we run a Montecarlo simulation of \eqref{eq:ITO} with $N=10^6$ cells and $dt=0.06$\,s. Then, we calibrated  $\sigma$  in order to obtain a stationary state with average $62.2^{\circ}$ and $\lambda_\theta$ to replicate the time evolution of data. In particular, we set $\sigma\approx 0.04 $ that is such that $\bar{\theta}_\ell  =62.2^{\circ}$ and $\lambda_\theta\approx 0.18$ \,s. After $1$ hour we have that the average orientation is $54.6^{\circ}$ and after $3$ hours the average orientation is $62.04^{\circ}$ and the $85\%$ of the cells is oriented at angles of $50^{\circ}$-$80^{\circ}$. In Fig. \ref{fig_Hayakawa} we plot both the histograms with classes' width of $10$ degrees and the recovered probability density functions (that are histograms with classes width of $0.01$ degrees.

\begin{figure}
\centering
\centering
\includegraphics[scale=0.5]{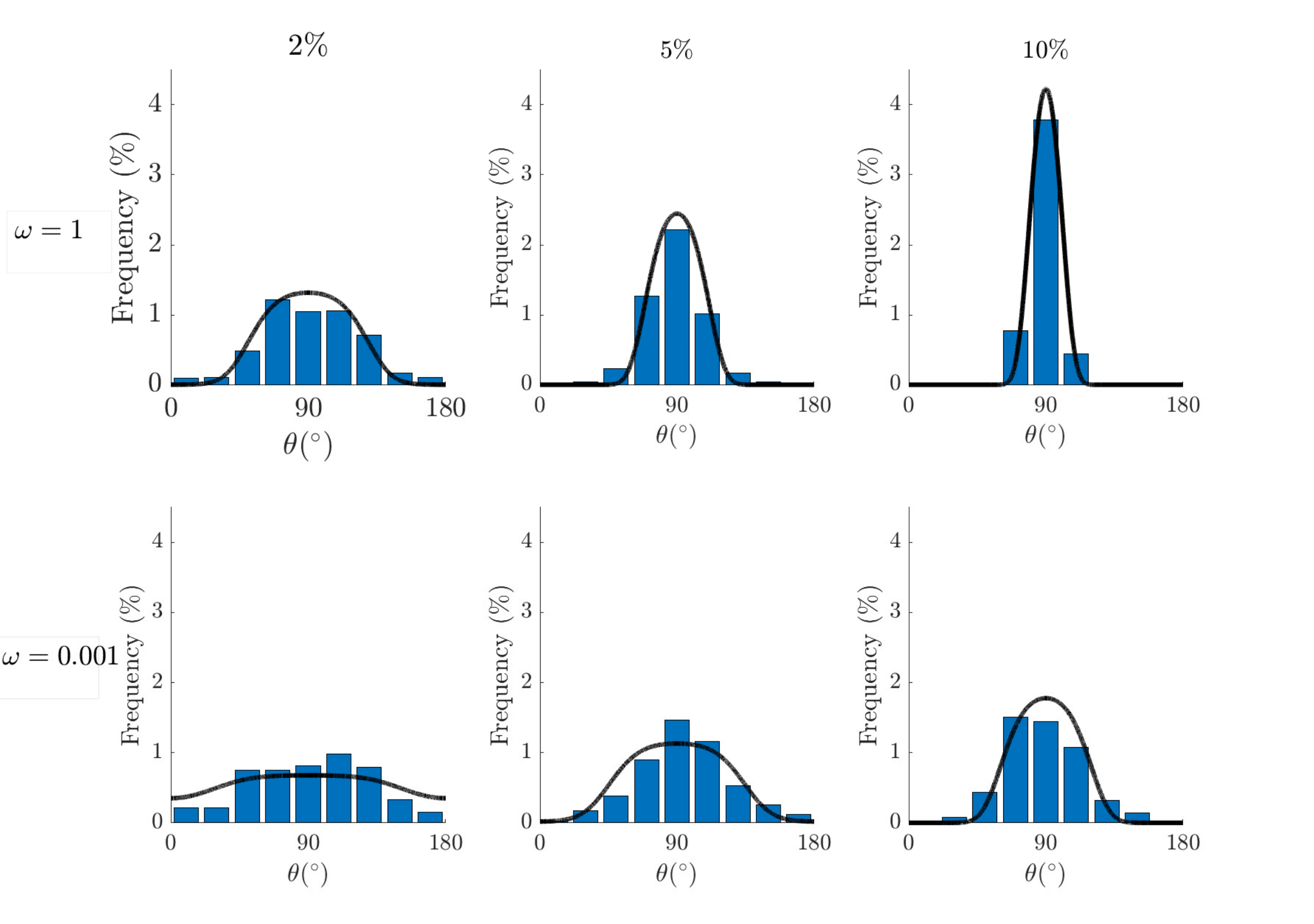}\\
\caption{Equilibrium distributions \eqref{eq:stat_state} changing $\bar\sigma$, $\omega$ and $\varepsilon$, while $\sigma=0.2$ everywhere. The parameter $\varepsilon=2\%,5\%,10\%$ (first, second and third columns, respectively) is changed according to the experimental setup giving rise to decreasing values of $\bar\sigma^2=\dfrac{\sigma^2}{2\varepsilon^2}$ in the first row and $\bar\sigma^2=\dfrac{\sigma^2}{2\varepsilon^2\lambda \omega}$ with $\lambda=100$\,s in the second row. The value of $\tilde{K}_s=0.7$ is used.}
\label{Mao1}
\end{figure}

Focusing on the stationary distributions, Mao et al. \cite{Mao} report some experimental data in histogram over $[0^\circ, 180^\circ)$ changing the stretching  amplitude ($\e=2\%,5\%,10\%$) and frequency ($\omega=1$ Hz, $0.001$ Hz). In particular, they show that increasing values of both amplitude and frequency lead to more peaked distributions. In their case, $r=0$ and the equilibrium orientation is perpendicular to the main stretching direction, i.e. $\theta_{eq}^1=90^\circ$. Trivially, due to symmetry, in this case mode and mean computed in $[0^\circ,180^\circ)$ coincide, with $\sigma, \varepsilon$ and $\omega$ determining only the variance of the probability density. In Fig. \ref{Mao1} in order to replicate the data reported by the histograms of \cite{Mao}, we plot \eqref{eq:stat_state} where we set the same $\sigma=0.2$ and vary $\e$ and $\omega$. When  $\omega=1$ (top row of Fig. \ref{Mao1}), that corresponds to a high frequency regime, increasing the strain amplitude, coherently with the fact that  $\bar{\sigma}^2=\dfrac{\sigma^2}{2\e^2}$ (so, it goes like $\e^{-2}$) we have more peaked distributions  that fit quite well the experimental distributions. 

For $\omega=0.001 \textrm{Hz}$ since $\lambda \omega$ corresponds to a low frequency regime (it is $\lambda\omega=0.1$ if we take $\lambda=100$\,s), we use $\bar{\sigma}^2=\dfrac{\sigma^2}{2\varepsilon^2\lambda \omega}$. 
Also in this case, the distributions peak up increasing the strain amplitudes. Also in this case the theoretical results compare well with  the experimental results, in spite of the fact that  we are not really using a viscoelastic model but only taking into account of viscoelastic effects through a modification of $\bar\sigma$ that is valid in the low frequency regime. 
Comparing the results obtained for a fixed $\e$ at the different $\omega$'s (for instance, the last column in Fig. \ref{Mao1}) simulations give more peaked distributions for higher frequencies.

\begin{figure}[!ht]
\centering
\includegraphics[scale=0.24]{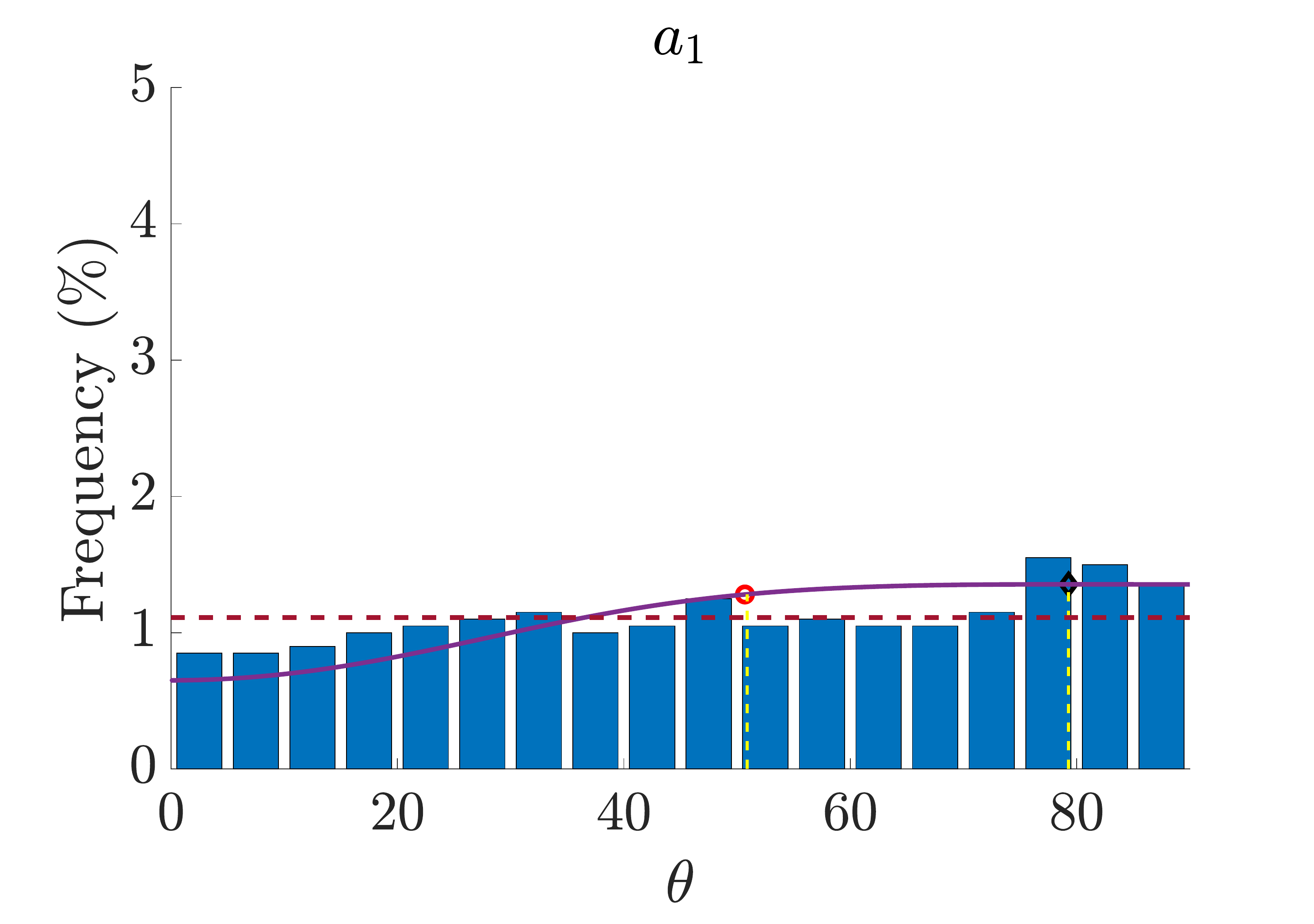}
\includegraphics[scale=0.24]{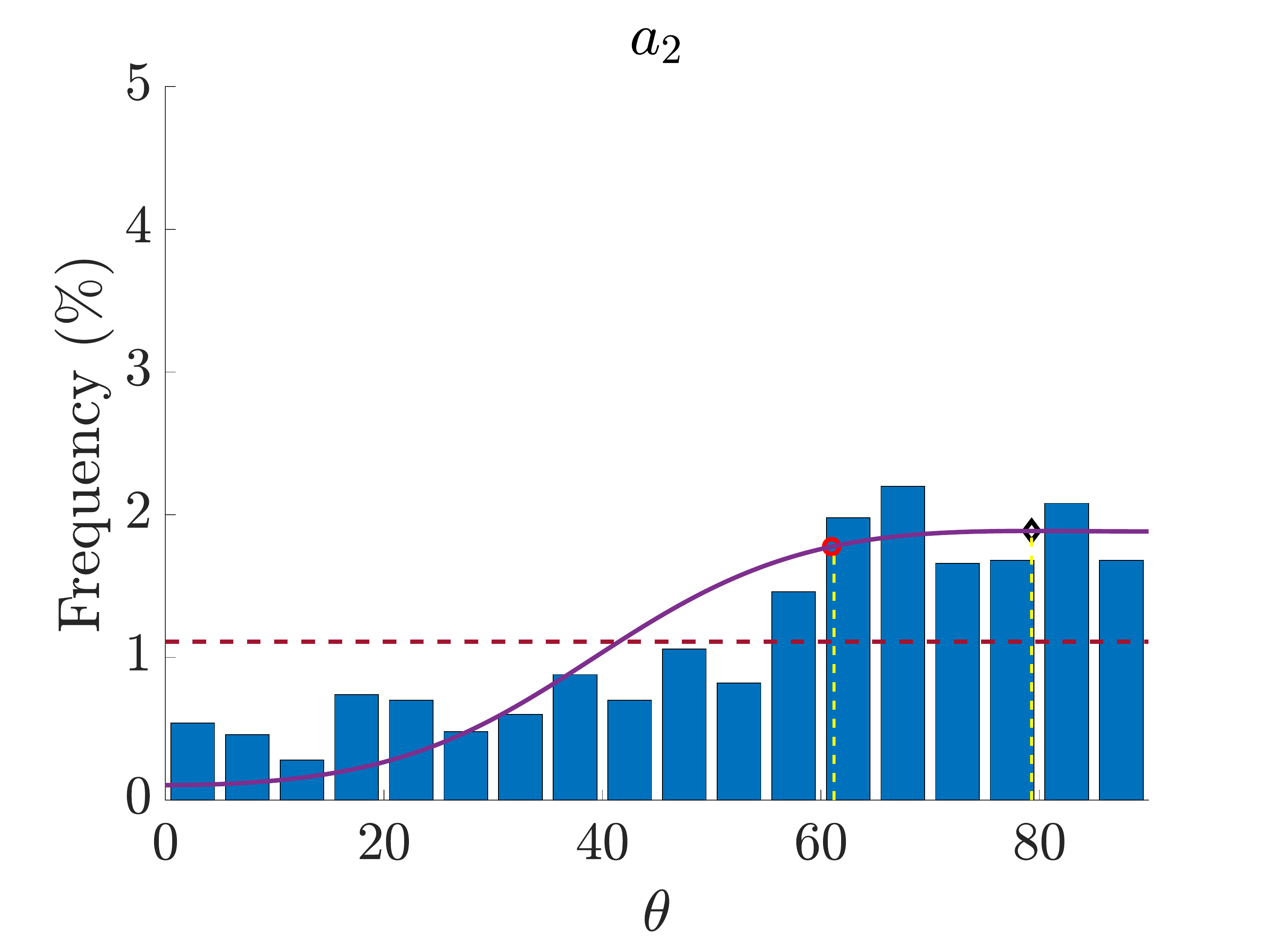}
\includegraphics[scale=0.24]{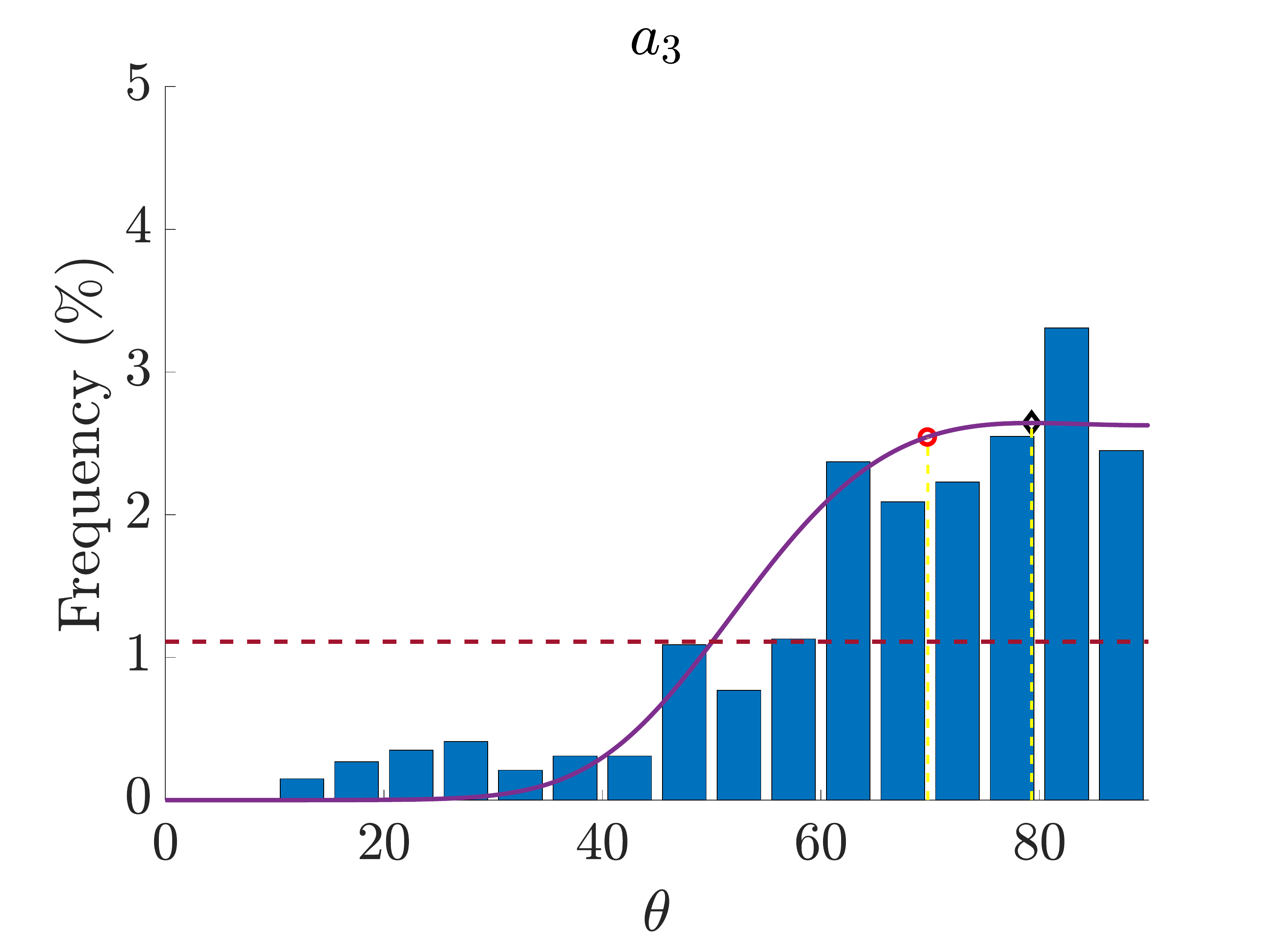}
\includegraphics[scale=0.24]{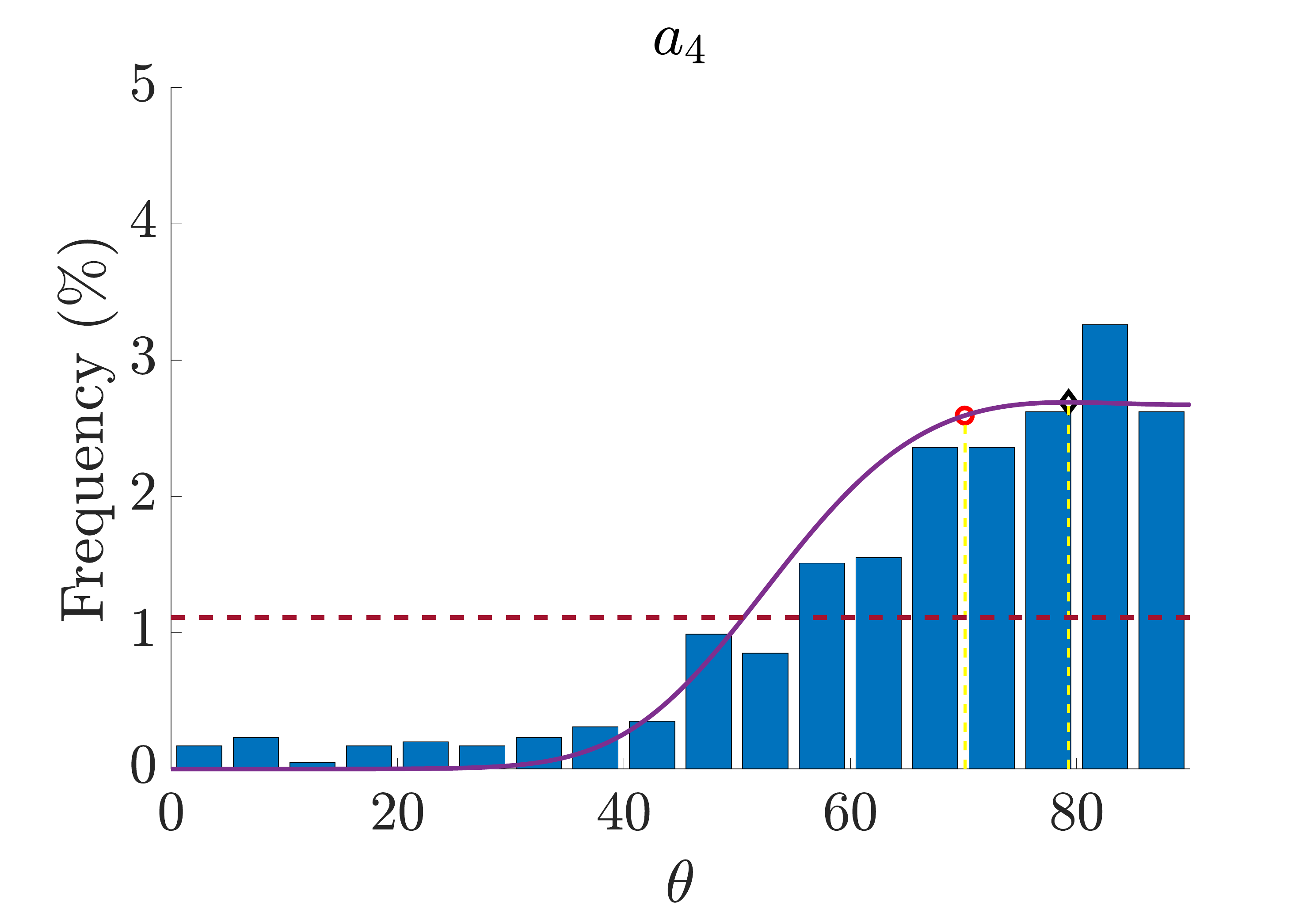}
\begin{tabular}{c|c|c|c|c}
\toprule
 &$a_1$ & $a_2$ & $a_3$& $a_4$\\
\hline
\hline
$\varepsilon (\%)$ & $4.9$ & $8.4$ & $11.8$ & $14$\\
$\bar{\theta}_\ell^{\textrm{hist}}$ & $51.1^\circ$ & $60.6^\circ$ & $70.01^\circ$ & $70.3^\circ$ \\
$\bar{sd}_\ell^{\textrm{hist}}$ & $26^\circ$ & $23^\circ$ & $17^\circ$ & $18^\circ$\\
$\bar{\theta}_\ell^{\infty}$ & $51.1^\circ$ & $60.6^\circ$ & $69.9^\circ$ & $70.2^\circ$ \\
$\sqrt{\bar{v}_\ell^{\infty}}$ & $24.4^\circ$ & $19.9^\circ$ & $12.9^\circ$ & $12.7^\circ$\\
$\sigma$ & $0.15$ & $0.2$ & $0.14$& $0.16$\\
\bottomrule
\end{tabular}
\caption{Equilibrium distributions  \eqref{eq:stat_state} with $\bar{\mathcal{U}}$ given by \eqref{barU}  in the cases $a_1,a_2,a_3,a_4$ reported in \cite{Faust} with applied strains listed in the table.  
In all figures we have $r=0.15$ and $\tilde{K}_s=0.7$ that allowed to best reproduce the averages of the histograms $\bar{\theta}_l^{\textrm{hist}}$ by varying $\sigma$ in \eqref{eq:stat_state}. The red circles represent the average circular orientation $\bar{\theta}_l^{\infty}$ computed using \eqref{eq:ave.c}. The black diamond represents $\theta_{eq}^1$.  We also computed the standard deviation of the histogram $\bar{sd}_\ell^{\textrm{hist}}$ and the standard deviation $\sqrt{\bar{v}_\ell^{\infty}}$ of the stationary state using \eqref{eq:var}.}  
\label{fig_Faust}
\end{figure}

Faust et al. \cite{Faust} report the results of some experiment characterized by an evaluated biaxiality ratio of $r=0.15$. Assuming that $\alpha=\alpha_L$, as also suggested in \cite{Livne}, the minimum elastic energy and therefore the mode is obtained at $\theta_{eq}^1\approx 79^\circ$.
They perform the experiment applying different stretching amplitudes, namely $4.9\%$ (denoted as Case $a_1$), $8.4\%$ (Case $a_2$), $11.8\%$ (Case $a_3$), and $14\%$ (Case $a_4$). 
We recall that in this case, at variance with the (symmetric) one in \cite{Mao}, the mean changes with the strain amplitude that influences $\bar\sigma$ (see second row in the table in Fig. \ref{fig_Faust}). The means of the stationary distribution obtained by the simulation reported in the fourth row in the table closely follow the experimental ones. A slight difference is found for the standard deviation, expecially for larger amplitudes.
Therefore, in Fig. \ref{fig_Faust} we compare their experimental results with the stationary probability density functions defined by \eqref{eq:stat_state} having average and standard deviation as computed from the histograms reported in \cite{Faust}. 

\section{Kinetic Description}
With the aim to get closer to the intrinsic dynamics followed by the single cell, in this section we will apply some classic tools of kinetic theory that, starting from the definition of the microscopic dynamics  performed by cells  to re-orient,  allow to derive the related mesoscopic evolution equation, such as  \eqref{FP0}. After going through the general procedure, we will then apply it to different microscopic rules. In particular, in Section \ref{sec:nonlocal} we will introduce a more realistic way cells may use  to non-locally sense the state of stress. This will lead to a non-local Fokker-Planck equation.  Then, in  Section \ref{sec:control} we will discuss a different intrinsic dynamics that is probably performed by the cell, that through an optimal control argument allows to align along the most convenient orientation.

\subsection{Derivation of kinetic models from discrete random processes}

As a first step we formalize a microscopic discrete random process for describing the reorientation of cells. Let $\Theta_t\in[0,\pi)$ denote a random variable describing the orientation of a representative cell at time $t$. 
As typically done in kinetic theory \cite{pareschi2013BOOK}, over a finite time interval $\Delta t$, we assume that  
a cell can change its main axis according to whether a re-orientation occurs or not. We then express this discrete-in-time random process as
\begin{equation}\label{eq:micro_discr}
\Theta_{t + \Delta t} = (1-T_{\lambda_\theta})\Theta_{t} + T_{\lambda_\theta}\Theta_t' ,
\end{equation}
where $\Theta_t'$ is the random variable in $[0,\pi)$ describing the new direction after a re-orientation given the previous direction $\Theta_t$, while $T_{\lambda_\theta}$ is a Bernoulli random variable which we assume to be independent of all the other variables
appearing in \eqref{eq:micro_discr}, discriminating whether the direction changes ($T_{\lambda_\theta}=1$) or not ($T_{\lambda_\theta}$=0) during the time interval $\Delta t$. In particular we set 
\begin{equation}\label{pdT}
Prob(T_{\lambda_\theta}=1)=\Delta t/\lambda_\theta ,
\end{equation} 
where the necessary condition for $T_{\lambda_\theta}$ to be a random variable is
\begin{equation}\label{lDt}
\Delta t/\lambda_\theta \le 1.
\end{equation}
The latter models our assumption according to which the larger the time interval is, the higher the probability of having a reorientation is. 
The quantity $\Theta_t'$ models the change of direction (if it happens) and it may be generally expressed as
\[
\Theta_t'=h_{\lambda,K}(\Theta_t)+\sqrt{\sigma^2}\xi \quad \textrm{mod}(\pi), 
\]
i.e. the new direction $\Theta_t'$  is a function $h_{\lambda,K}$ of the previous orientation $\Theta_t$ and of the deformation parameters $\lambda_x,\lambda_y,K_{\|},K_{\bot}, K_s$,  accounted for by the index $\lambda,K$. 
We shall assume $h_{\lambda,K}$ to be a regular function of its arguments, i.e. $h_{\lambda,K} \in \mathcal{C}^1([0,\pi))$, $\xi$  is a standard gaussian random variable, i.e. $\xi \sim \mathcal{N}(0,1)$ satisfying $\ave{\xi}=0$, $\ave{\xi^2}=1$, while the term $\textrm{mod}(\pi)$ models the fact that $\Theta_t$ is $\pi$-periodic.

We now want to recover an aggregate description of the orientations of the cells in order to obtain a statistical description of the orientations themselves. Let  then $\varphi=\varphi(\theta)$ be an observable quantity defined on the phase space $[0,\pi)$. From~\eqref{eq:micro_discr} together with the assumed independence of $T_{\lambda_\theta}$, we see that the mean variation rate of $\varphi$ in the time interval $\Delta{t}$ satisfies
\begin{eqnarray*}
	\frac{\ave{\varphi\left(\Theta_{t+\Delta t}\right)}-\ave{\varphi\left(\Theta_{t}\right)}}{\Delta{t}}
& = &\frac{\ave{\varphi\left((1-T_{\lambda_\theta})\Theta_{t} + T_{\lambda_\theta}\Theta_t'\right)}-\ave{\varphi\left(\Theta_{t}\right)}}{\Delta{t}}\\
& =	&\frac{\ave{\varphi\left(\Theta_{t}\right)}(1-\Delta t/\lambda_\theta) + \ave{\varphi\left(\Theta_t'\right)}\Delta t/\lambda_\theta-\ave{\varphi\left(\Theta_{t}\right)}}{\Delta{t}},
\end{eqnarray*}
where here and henceforth $\ave{\Psi}$ denotes the expectation of a generic random variable $\Psi$ with respect to its law. Then, the latter equality holds remembering that $\ave{T_{\lambda_\theta}}=\Delta t/\lambda_\theta$ and that $T_{\lambda_\theta}$ is independent from all the other random variables.
Whence, we deduce the instantaneous time variation of the average of $\varphi$ in the limit $\Delta{t}\to 0^+$ as
$$ \frac{d}{dt}\ave{\varphi\left(\Theta_t\right)}=\dfrac{1}{\lambda_\theta}\ave{\big(\varphi\left(\Theta_t'\right)-\varphi\left(\Theta_t)\right)}.
 $$
 If $f(t,\theta)$ is a probability density function,
then we obtain
\begin{equation}\label{eq:Boltz}
\begin{split}
\frac{d}{dt}\int_0^{\pi}\varphi(\theta)f(t,\,\theta)\,d\theta=
\dfrac{1}{\lambda_\theta}\ave{\int_0^{\pi}\left(\varphi(\theta')-\varphi(\theta) \right)f(t,\, \theta) d\theta} ,
\end{split}
\end{equation} 
where $\theta'$ is given by
\begin{equation}\label{eq:micro_gen}
\theta'=h_{\lambda,K}(\theta)+ \sqrt{\sigma^2}\xi \quad \textrm{mod}(\pi).
\end{equation} 
 Equation \eqref{eq:Boltz} is a Boltzmann-type integro-differential equation.

Choosing $\varphi(\theta)=1$ we readily obtain
$$ \frac{d}{dt}\int_0^{\pi} f(t, \theta)\,d\theta=0, $$
which means that the total mass of the agents is conserved in time by the interactions~\eqref{eq:micro_gen}.
Classically, the evolution of the statistical moments of $f$ are obtained choosing $\varphi(\theta)=\theta^n$, $n=1,\,2,\,\dots$.

\subsection{Quasi-invariant direction limit}\label{sec:quasi}
One of the most relevant aspects of kinetic models is the
possibility of characterising the stationary distributions arising asymptotically for $t\to +\infty$. This is typically carried out by means of asymptotic procedures, which, in suitable regimes of the parameters of the microscopic interactions, allows to transform a Boltzmann-type integro-differential equation into a partial differential equation usually easier to be investigated analyticallly. A particularly efficient asymptotic procedure is the so-called \textit{quasi-invariant limit}, which leads to \textit{Fokker-Planck-type} equations.

The idea behind the quasi-invariant limit is that one studies a regime in which the new reorientation direction $\theta'$ is close enough to the previous direction $\theta$, so that the reorientation enhances a small variation. This concept was first introduced in the kinetic literature on multi-agent systems in~\cite{cordier2005JSP,toscani2006CMS} for binary collisions and in~\cite{furioli2017M3AS} for the interactions with a fixed background and has its roots in the concept of \textit{grazing collisions} studied in the classical kinetic theory~\cite{villani1998ARMA}.

In our framework this corresponds to considering a small re-orientation and, then, a rescaled microscopic rule \eqref{eq:micro_gen}
\begin{equation}
	\theta'=\theta+\gamma \left(h_{\lambda,K}(\theta)-\theta\right)+\sqrt{\gamma {\sigma^2}}\xi, \quad \textrm{mod}(\pi), 
	\label{eq:quasi-invariant_scaling}
\end{equation}
where $\gamma \ll 1$.  Now, diffusion is linked to a random variable $\sqrt{\gamma}\xi=\mathcal{N}(0,\gamma)$ with zero mean and variance $\gamma$.

To compensate for the smallness of each re-orientation, we simultaneously scale time as $\tau:=\gamma t$,
which corresponds to observe the dynamics on a slower time scale and we introduce
\[
f^{\gamma}(\tau,\theta)=f(\tau/\gamma,\theta).
\] 
Equivalently we can scale $\lambda_\theta$ as $\lambda_\theta^\gamma:=\gamma \lambda_\theta$, meaning that the re-orientation time corresponding to a small re-orientation is shorter when $\gamma\ll 1$.
Then, Eq. \eqref{eq:Boltz} rewrites
\begin{equation}\label{eq:Boltz_resc}
\frac{d}{d\tau}\int_0^{\pi}\varphi(\theta)f^{\gamma}(\tau,\theta)\,d\theta=\dfrac{1}{\gamma\lambda_\theta}\ave{\int_0^{\pi}\left(\varphi(\theta')-\varphi(\theta) \right)f^{\gamma}(\tau,\theta) d\theta} ,
\end{equation}
with \eqref{eq:quasi-invariant_scaling}.
Now, let $\varphi \in \mathcal{C}^3([0,2\pi))$ satisfy the requirement $\varphi(0)=\varphi(\pi)=0$, as we are considering quasi-invariant changes of a direction belonging to $[0,\pi)$ \cite{FesTosWol2018KRM}. Expanding the difference $\varphi(\theta')-\varphi(\theta)$ in Taylor series about $\theta$ and using~\eqref{eq:quasi-invariant_scaling} we get
\begin{align}
	\begin{aligned}[b]
		\frac{d}{d\tau}\int_0^{\pi}\varphi(\theta)f^{\gamma}(\tau,\theta)\,d\theta &= \dfrac{1}{\gamma\lambda_\theta}\int_0^{\pi}\dfrac{d\varphi}{d\theta}(\theta)\gamma (h_{\lambda,K}(\theta)-\theta)f^{\gamma}(\tau,\theta) \, d\theta\\
		&\phantom{=} +\frac{1}{2\gamma\lambda_\theta}\int_0^{\pi}\dfrac{d^2\varphi}{d\theta^2}(\theta)\gamma \sigma^2 f^{\gamma}(\tau,\theta)\,d\theta+R_\varphi(f^{\gamma})(\tau,\theta),
	\end{aligned}
	\label{eq:FP+R}
\end{align}
where 
\[
R_\varphi(f^{\gamma})(\tau,\theta)= \dfrac{1}{\gamma\lambda_\theta}\int_0^{\pi}\dfrac{1}{2}\gamma^2
\dfrac{d^2\varphi}{d\theta^2}(\theta) (h_{\lambda,K}(\theta)-\theta)\, d\theta +
\dfrac{1}{\gamma\lambda_\theta}\int_0^{\pi}\dfrac{1}{6}
\dfrac{d^3\varphi}{d\theta^3}(\theta+\delta(\theta'-\theta)) \ave{(\theta'-\theta)^3}\, d\theta ,
\] 
being $\delta \in (0,1)$. If we assume that $|\xi^3|<\infty$ the following holds\footnote{Here and henceforth we use the notation $a\lesssim b$ to mean that there exists a constant $C>0$, independent of $\gamma$ and whose specific value is unimportant, such that $a\leq Cb$.} (cf.~\cite{cordier2005JSP} for similar calculations)
\begin{align}
	\begin{aligned}[b]
		\abs{R_\varphi(f^{\gamma})(\tau,\theta)} &\lesssim \left\Vert\dfrac{d^2\varphi}{d\theta^2}\right\Vert_\infty
\gamma\int_0^{\pi}(h_{\lambda,K}(\theta)-\theta)^2f^{\gamma}(\tau,\theta)\,d\theta \\
		&\phantom{\lesssim} +\left\Vert\dfrac{d^3\varphi}{d\theta^3}\right\Vert_\infty
\int_0^{\pi}\left(\gamma^2(h_{\lambda,K}(\theta)-\theta)^3+\sqrt{\gamma}4\sigma^3+3\gamma\sigma^2(h_{\lambda,K}(\theta)-\theta)^3\right)f^{\gamma}(\tau,\theta)\,d\theta.
	\end{aligned}
	\label{eq:remainder}
\end{align}
If we assume that $h^2_{\lambda,K}, h^3_{\lambda,K}$ are bounded in $[0,\pi)$, as $F2$ is satisfied, then
$$ R_\varphi(f^\gamma)\xrightarrow{\gamma\to 0^+}0. $$
Let us assume now that $(f^\gamma)$ converges in $C([0,\pi);\,L^1([0,\pi))\cap L^1([0,\pi);\,\theta\,d\theta))$, possibly up to subsequences, to a distribution function $f$ when $\gamma\to 0^+$.  Then, passing to the limit $\gamma\to 0^+$ in~\eqref{eq:FP+R} we obtain the limit equation
\begin{align*}
	\frac{d}{dt}\int_0^{\pi}\varphi(\theta)f(\tau,\theta)\,d\theta &= 
	\dfrac{1}{\lambda_\theta}\int_0^{\pi}\dfrac{d\varphi}{d\theta}(\theta)(h_{\lambda,K}(\theta)-\theta)f(\tau,\theta)\, d\theta \\
	&\phantom{=} +\frac{1}{2\lambda_\theta}\int_0^{\pi}\dfrac{d^2\varphi}{d\theta^2}(\theta)\sigma^2f(\tau,\theta)\,d\theta,
\end{align*}
which, by integration by parts and recalling the compactness of the support of $\varphi$, can be recognised as a weak form of the following Fokker-Planck equation 
\[
\dfrac{\partial}{\partial \tau}f(\tau,\theta)=-\dfrac{1}{\lambda_\theta}\dfrac{\partial}{\partial \theta}\left[(h_{\lambda,K}(\theta) -\theta)f(\tau,\theta)\right]+\dfrac{1}{2\lambda_\theta}\dfrac{\partial^2 }{\partial \theta^2}\left[\sigma^2 f(\tau,\theta)\right].
\]

\subsection{Microscopic Evolution of the Cell Orientation}
We shall now study the result of introducing specific microscopic rules describing cell re-orientation.

\subsubsection{Local and non-local evaluation of elastic energy}\label{sec:nonlocal} 

If we want to model the new orientation of a cell that tries to minimize a potential energy $\mathcal{U}$ after a time interval $d t$ we may observe that the discrete in time random process describing the evolution of the orientation $\Theta_t$ happens with frequency $1/\lambda_\theta$ and may be expressed by discretizing \eqref{eq:ITO} over $d t$ (where we consider the high frequencies regime) and setting $d t=\gamma$

\begin{equation}\label{eq:model1_inv}
\theta'=\theta -\gamma \varepsilon^2\dfrac{\partial \bar{\mathcal{U}}}{\partial \theta}+\sqrt{\gamma\sigma^2}\xi\quad \textrm{mod}(\pi).
\end{equation}
Using the results obtained in the previous section, we have the Fokker-Planck equation 
\begin{equation}\label{FP}
\dfrac{\partial}{\partial \tau}f(\tau,\theta)=\dfrac{\varepsilon^2}{\lambda_\theta}\dfrac{\partial}{\partial \theta}\left( \dfrac{\partial \bar{\mathcal{U}}}{\partial \theta}f(\tau,\theta)\right)+\dfrac{1}{2\lambda_\theta}\dfrac{\partial^2 }{\partial \theta^2}\left[\sigma^2 f(\tau,\theta)\right],
\end{equation}
which is, as expected, the same as \eqref{FP0}.

Actually, rather than the local rule above,
we can start from the observation that
at the sub-cellular level, cells sense nonlocally the environment, in this case  the energy field, and actively respond to the cues. So, we can assume that rather than a local gradient they non-locally measure
\begin{equation}\label{eq:non_local_flux}
h_{\lambda,K}(\theta)=\theta-\dfrac{1}{\theta_0^2}\displaystyle \int_{-\theta_0}^{\theta_0} k(|\beta|)\varepsilon^2\bar{\mathcal{U}}(\theta+\beta)\dfrac{\beta}{|\beta|}\, d\beta\,,
\end{equation}
i.e., the cell averages the energy in the angles spanning the interval $[-\theta_0,\theta_0]$ weighting it according to the sensing kernel $k$ (see \cite{Armstrong_Painter_Sherratt.06,loy2019JMB, Loy_Preziosi2}).
In this case one gets the non-local Fokker-Planck equation
\[
\dfrac{\partial}{\partial \tau}f(t,\theta)=\dfrac{1}{\lambda_\theta}\dfrac{\partial}{\partial \theta}\left[\left( 
\dfrac{1}{\theta_0^2}\displaystyle \int_{-\theta_0}^{\theta_0} k(|\beta|)\varepsilon^2\bar{\mathcal{U}}(\theta+\beta)\dfrac{\beta}{|\beta|}\, d\beta\right) f(\tau,\theta)\right]+\dfrac{1}{2\lambda_\theta}\dfrac{\partial^2 }{\partial \theta^2}\left[\sigma^2 f(\tau,\theta)\right].
\]

Equation \eqref{FP} can then be recovered in the limit $\theta_0\to 0$ if $k$ and $\bar{\mathcal{U}}$ are differentiable. 

\subsubsection{Re-orientation as an optimal  control problem}\label{sec:control}
In this section we want to introduce a new point of view consisting in modelling re-orientation as a result of an internal control actuated by the cell. From the mathematical point of view, this approach consists in expressing re-orientation rules like \eqref{eq:micro_gen} starting from a control problem, in the sense that we assume that the cell changes its reorientation by an angle $\nu\psi_{opt}$ where $\psi_{opt}$ is the angle that minimizes a certain cost functional $\mathcal{J}$.
Therefore, in the same spirit as for instance \cite{preziosi2021JTB}, we write
\begin{equation}\label{eq:model2_cont}
\theta'=\theta+\nu \psi_{opt}, \quad \psi_{opt}=\textrm{argmin}_\psi \mathcal{J}(\psi),
\end{equation}
where 
\[
\mathcal{J}(\psi)=\nu\dfrac{\psi^2}{2}+\langle g(\theta')\rangle,
\]
and the function $g$ will be specialized later on.

In order to determine the optimal control  at each orientation, we need to introduce a Lagrangian
\begin{equation}\label{eq:Lagrangian}
\mathcal{L}(\theta',\psi)=\mathcal{J}(\psi)+\eta \ave{\theta'-(\theta+\nu \psi)},
\end{equation}
where $\eta \in \mathbb{R}$ is the Lagrange multiplier associated with the constraint \eqref{eq:model2_cont}.
The optimality conditions are then identified by the solution of 
\begin{equation}
\left\{\begin{aligned}
\dfrac{\partial \mathcal{L}}{\partial\theta'}&=\left\langle \dfrac{dg}{d\theta'}(\theta')\right\rangle+\eta=0\\
\dfrac{\partial \mathcal{L}}{\partial\psi}&=\nu(\psi -\eta)=0
\end{aligned}
\right.
\end{equation}
So, eliminating the Lagrange multiplier, the optimal value is implicitly identified by
\begin{equation}\label{eq:opt_cont}
\psi_{opt}+\left\langle \dfrac{dg}{d\theta'}(\theta')\Big |_{\theta'=\theta+\nu\psi_{opt}}\right\rangle=0\,.
\end{equation}
If we choose $g=\e^2\bar{\mathcal{U}}$, then Eq.  \eqref{eq:opt_cont} specializes to
$$\psi_{opt}+\e^2\dfrac{d\bar{\mathcal{U}}}{d\theta}(\theta+\nu\psi_{opt})=0\,,$$
that, in general, allows to determine the optimal control only implicitly.

In any case the re-orientation rule \eqref{eq:model2_cont} specializes into
$$\theta'=\theta+\nu\psi_{opt}=\theta-\nu \e^2\dfrac{d\bar{\mathcal{U}}}{d\theta}(\theta+\nu\psi_{opt})\,,$$
that in the limit of small $\nu$ used for the grazing limit and adding the stochastic term is equivalent to \eqref{eq:model1_inv} and leads again to \eqref{FP}.


In order to explicitly determine the control, we can instead more classically take a quadratic form for $g$ 
$$g(\theta')=\dfrac{\e^2}{2}[\theta'-\hat\theta(\theta)]^2\,,$$
where, assuming to work in Case 1,  
\[
\hat\theta(\theta)=\theta_{eq}^1p(\theta)+(1-p(\theta))(\pi-\theta_{eq}^1)
\] 
with $p(\theta)$  a non negative and continuous function defined on $[0,\pi)$  that satisfies
\[
p(\theta_{eq}^1)=1 \quad p(\pi-\theta_{eq}^1)=0,\quad p(0)=p(\pi/2)=p(\pi)=1/2,\quad 
p'(\theta_{eq}^1)=p'(\pi-\theta_{eq}^1)=0
\]
so that
\begin{equation}\label{thetahat}
\hat\theta(\theta_{eq}^1)=\theta_{eq}^1\qquad {\rm and}\qquad \hat\theta(\theta_{eq}^2)=\theta_{eq}^2\,.
\end{equation}
In this case one can explicitly solve \eqref{eq:opt_cont} and determine 
$$\psi_{opt}=-\,\dfrac{\varepsilon^2}{1+\nu \varepsilon^2}(\theta-\hat\theta)\,,$$
and therefore the re-orientation rule \eqref{eq:model2_cont} becomes
\begin{equation}\label{eq:model2_int}
\theta'=\theta+\gamma \e^2(\hat\theta(\theta)-\theta)\qquad{\rm where}\quad 
\gamma=\dfrac{\nu}{1+\nu \varepsilon^2}\,.
\end{equation}
Adding a stochastic fluctuation weighted by $\sigma_c$ we have
\begin{equation}\label{eq:model2}
\theta'=\theta+\gamma\e^2(\hat\theta(\theta)-\theta) +\sqrt{\gamma \sigma_c^2}\xi \quad \textrm{mod} (\pi).
\end{equation}
This rule implies the fact that at each re-orientation the cell  will activate a control to reach a better orientation that is given by a rotation of $\gamma\e^2(\hat\theta(\theta)-\theta)$ (plus a white noise). This process will stop when the cell has oriented along the stable equilibria, because of \eqref{thetahat}. 
In the symmetry points $\theta=0,\pi/2,\pi$ the cell has the same probability ($=1/2$) of re-orienting towards $\theta_{eq}^1$ or $\theta_{eq}^2=\pi-\theta_{eq}^1$. 


As illustrated in Section \ref{sec:quasi} in this case the quasi-invariant direction limit procedure leads to the following Fokker-Planck equation
\begin{equation}\label{eq:FP_model2}
\dfrac{\partial}{\partial \tau}f(\tau,\theta)=
-\,\dfrac{\e^2}{\lambda_\theta}\dfrac{\partial}{\partial \theta}\left[ (\hat\theta(\theta)-\theta)f(\tau,\theta)\right]
+\dfrac{1}{2\lambda_\theta}\dfrac{\partial^2}{\partial \theta^2}\left[\sigma_c^2 f(\tau,\theta)\right],
\end{equation}
that can be coupled with boundary conditions $F3$.

Therefore, the stationary state is given by
\[
- (\hat{\theta}(\theta)-\theta)f^\infty(\theta)+\dfrac{\partial }{\partial \theta }\left[\bar{\sigma}_c^2f^\infty(\theta)\right]=0\,,
\]
where $\bar{\sigma}_c^2=\dfrac{\sigma_c^2}{2\e^2}$ that gives
\begin{equation}\label{eq:stat_mod2}
f^{\infty}(\theta)=C\exp \left( \int_0^{\theta} \dfrac{\hat\theta(\theta)-\theta}{\bar{\sigma}_c^2} d\theta\right)
\end{equation}
where $C$ is the normalization constant.
This distribution has actually mode $\theta_{eq}^1$ and $\pi-\theta_{eq}^1$ in $[0,\pi)$ and average depending on the value of $\sigma_c$.



\begin{figure}[!ht]
\centering
\includegraphics[scale=0.24]{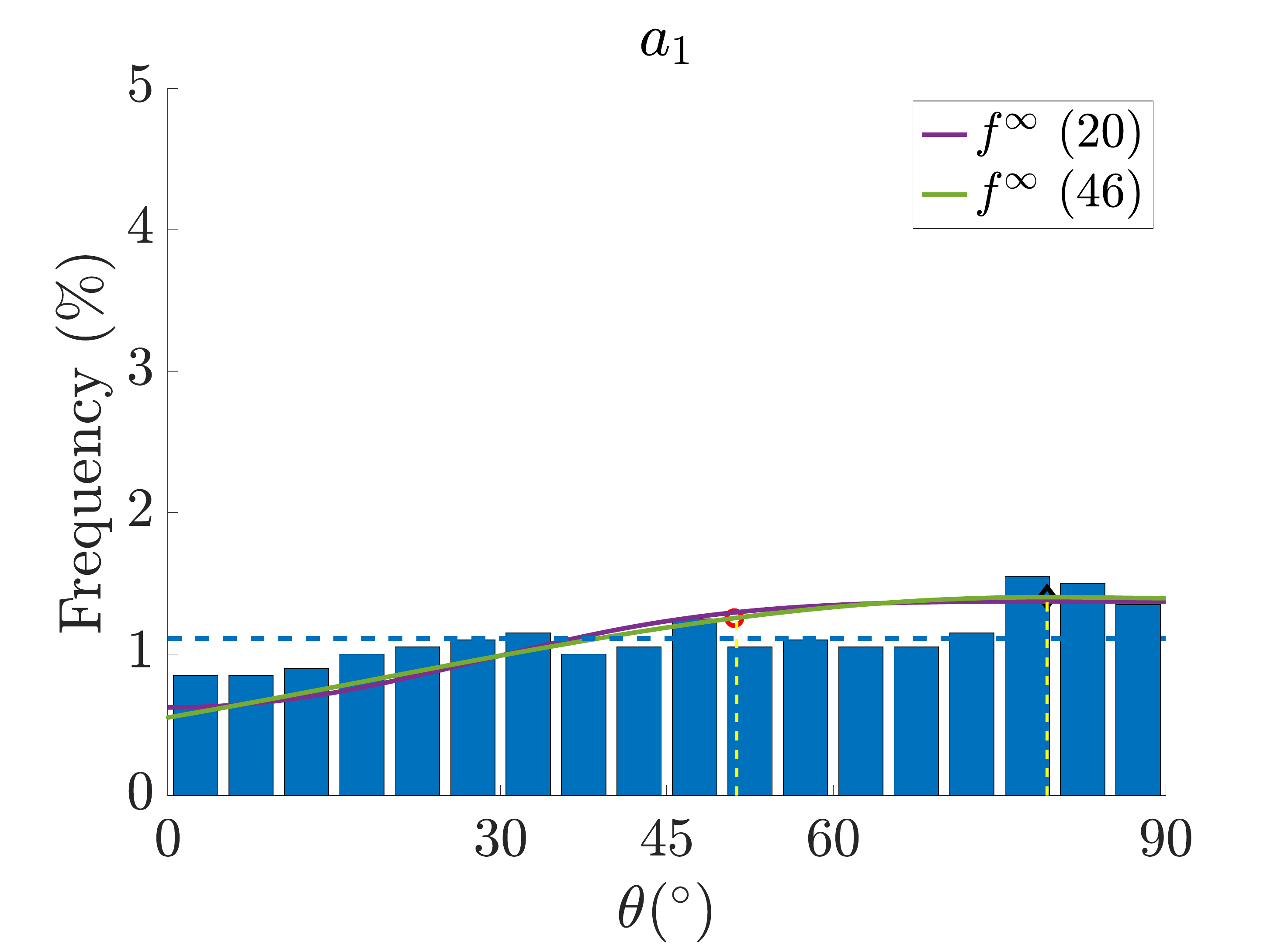}
\includegraphics[scale=0.24]{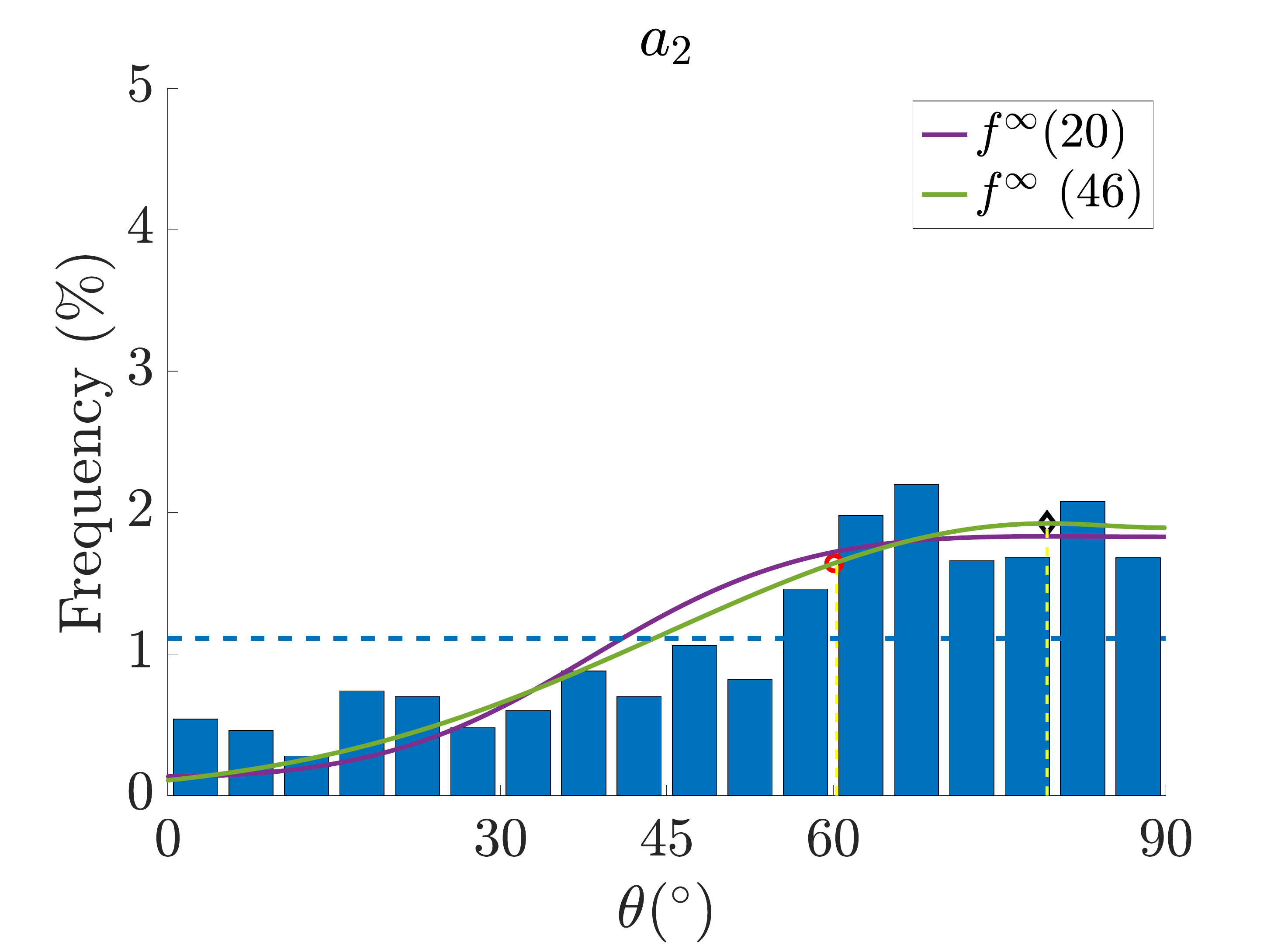}
\includegraphics[scale=0.24]{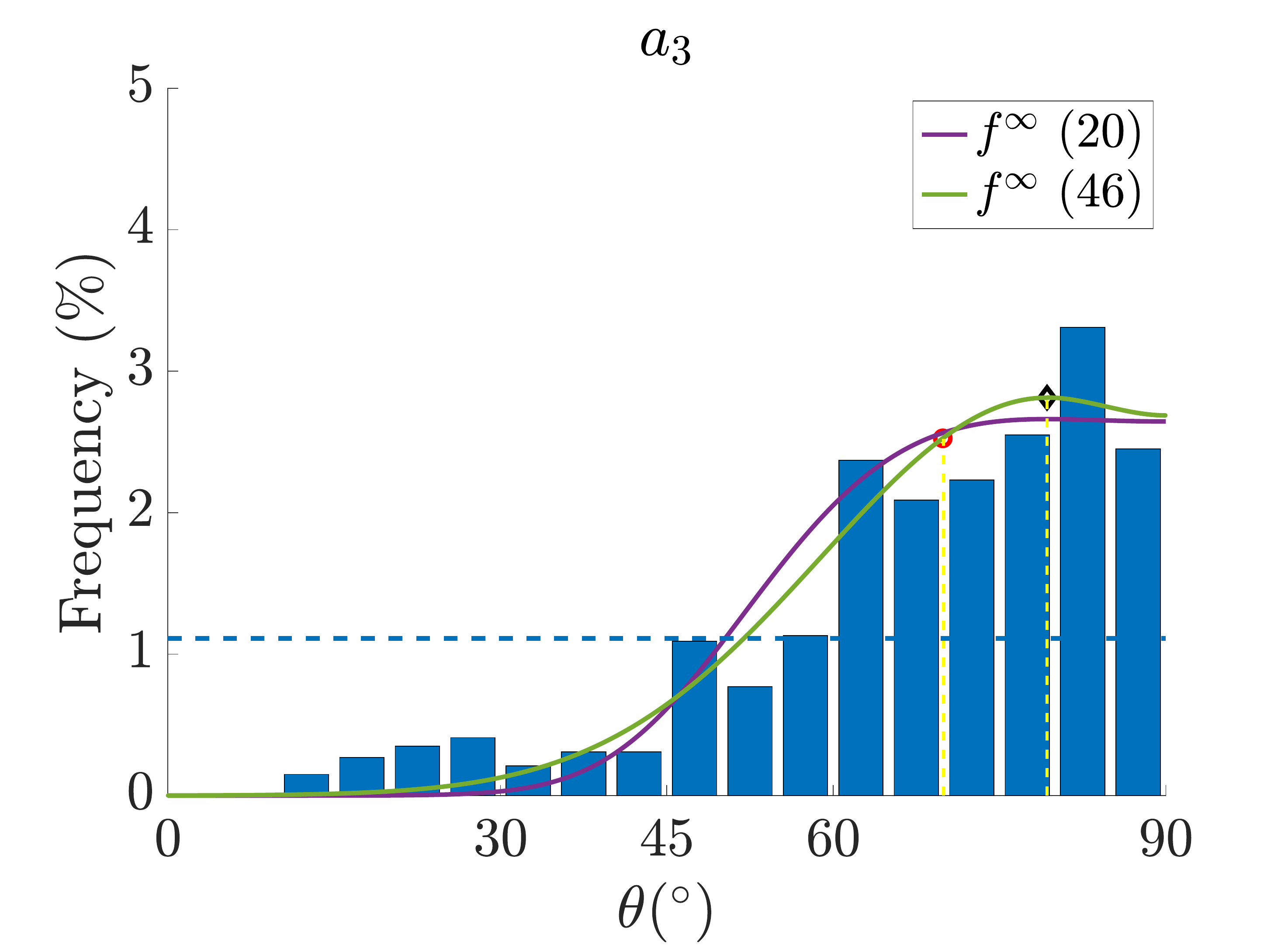}
\includegraphics[scale=0.24]{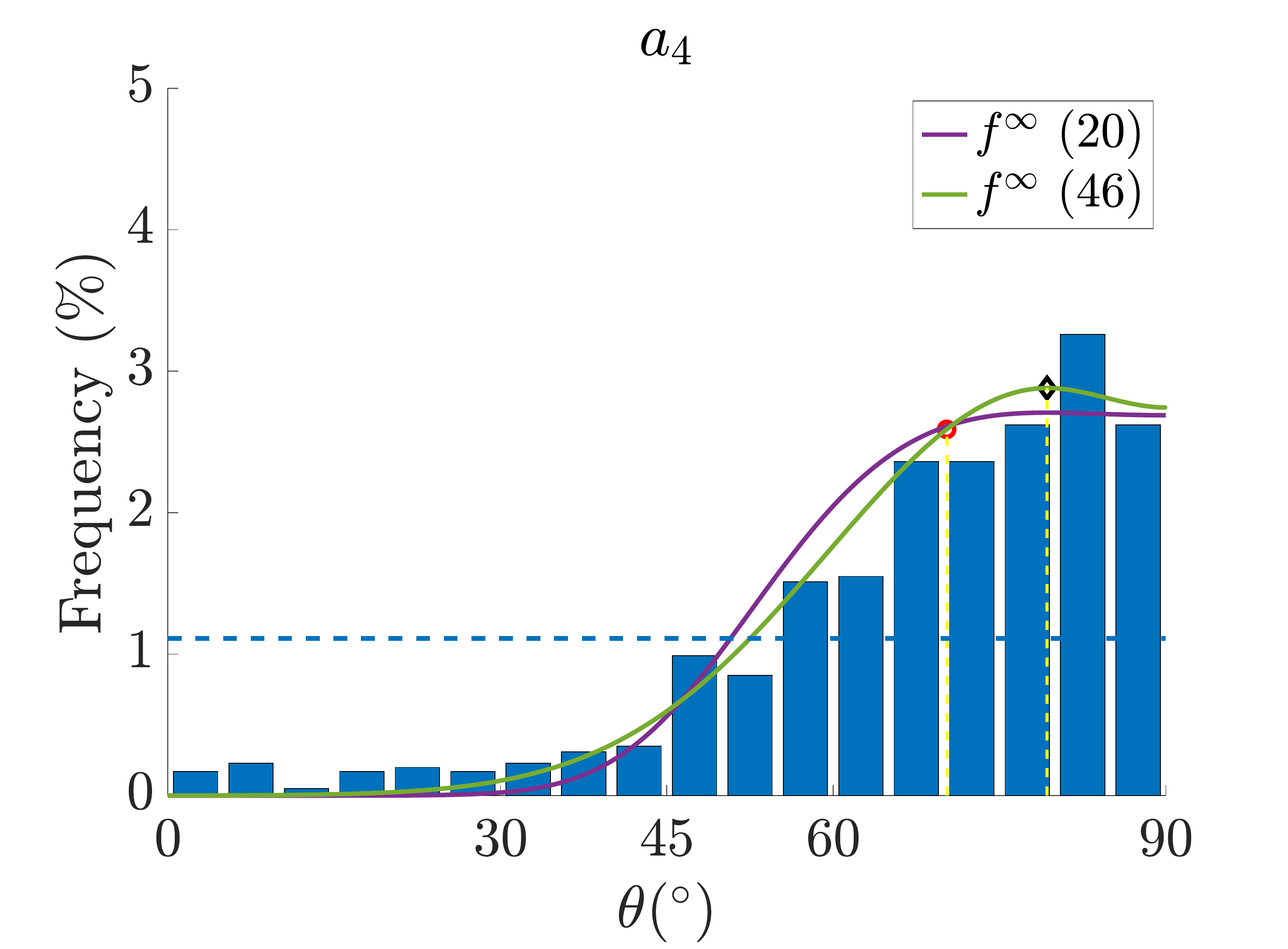}\\
\begin{tabular}{c|c|c|c|c}
\toprule
 &$a_1$ & $a_2$ & $a_3$& $a_4$\\
\hline
\hline
$\varepsilon (\%)$ & $4.9$ & $8.4$ & $11.8$ & $14$\\
$\bar{\theta}_\ell^{\textrm{hist}}$ & $51.1^\circ$ & $60.1^\circ$ & $70^\circ$ & $70.3^\circ$ \\
$\bar{sd}_\ell^{\textrm{hist}}$ & $26^\circ$ & $23^\circ$ & $17^\circ$ & $18^\circ$\\
$\bar{\theta}_\ell^{\infty}$ & $51.1^\circ$ & $60.1^\circ$ & $70^\circ$ & $70.3^\circ$ \\
$\sqrt{\bar{v}_\ell^{\infty}}$ & $24.6^\circ$ & $20.6^\circ$ & $13.8^\circ$ & $13.4^\circ$\\
$\sigma_c$ & $4.25$ & $4.11$ & $3.34$& $3.85$\\
\bottomrule
\end{tabular}

\caption{Equilibrium distributions \eqref{eq:stat_mod2} with $\bar{\sigma}^2=\dfrac{\sigma^2}{2\e^2}$ in the cases $a_1,a_2,a_3,a_4$ as listed in the table.  
As in Fig. \ref{fig_Faust}, in all figures we have $r=0.15$ and $\tilde{K}_s=0.7$ that allowed to best reproduce the averages of the histograms $\bar{\theta}_l^{\textrm{hist}}$ by varying $\sigma_c$ in \eqref{eq:stat_mod2}. The red circles represent the average circular orientation $\bar{\theta}_l^{\infty}$ computed using \eqref{eq:ave.c} with \eqref{eq:stat_mod2}. The black diamonds represent $\theta_{eq}^1$.  We also computed the standard deviation of the histogram $\bar{sd}_\ell^{\textrm{hist}}$ and the standard deviation $\sqrt{\bar{v}_\ell^{\infty}}$ of the stationary state using \eqref{eq:var} with \eqref{eq:stat_mod2}. We also superpose \eqref{eq:stat_state} with $\bar{\mathcal{U}}$ given by \eqref{barU} as reported in Fig. \ref{fig_Faust}. } 
\label{fig_Faust_2}
\end{figure}

In Fig. \ref{fig_Faust_2} we compare the stationary distribution \eqref{eq:stat_mod2} with the experimental data by Faust et al. \cite{Faust}, as in  Fig. \ref{fig_Faust}. Setting $\sigma_c$ in such a way that $\bar{\theta}_l$ of \eqref{eq:stat_mod2} is the same as in \cite{Faust}, we find that the microscopic rule \eqref{eq:model2} allows to recover a probability density function even better distributions than those in Fig. \ref{fig_Faust}. 
The prediction of the standard deviation in the fourth line of the table in the two figures show that those of \eqref{eq:stat_mod2} are slightly closer to the linear standard deviation reported in \cite{Faust}. We remark that the values of $\sigma$ and $\sigma_c$ are very different, and this is due to the fact that the rule \eqref{eq:ITO} expresses the variation of $\theta$ in terms of its derivative, while \eqref{eq:model2} expresses the variation through a rotation angle that the cell performs during a re-orientation.

\begin{figure}[!htbp]
\begin{center}
\subfigure[]{\includegraphics[scale=0.3]{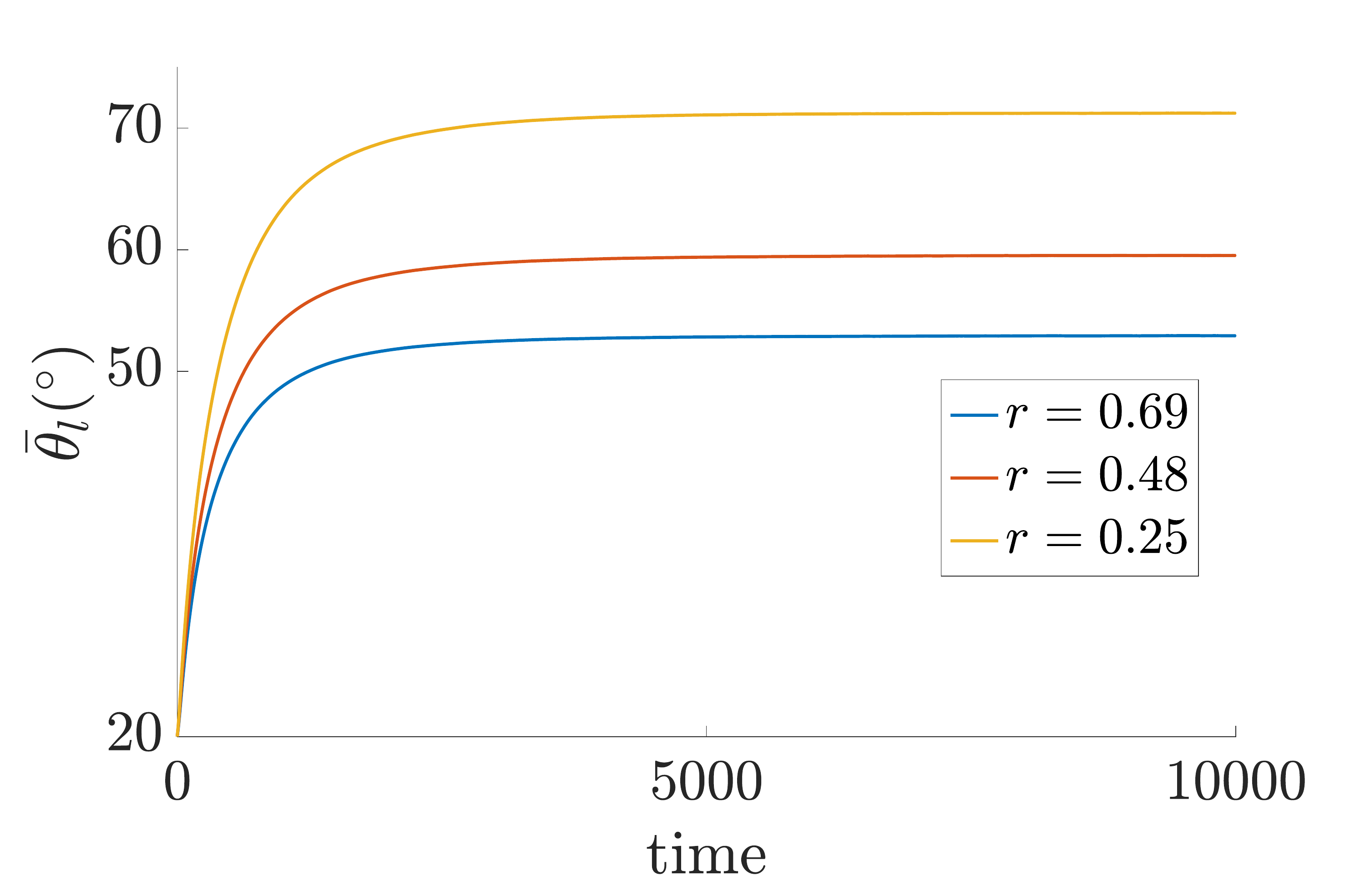}}
\end{center}
\begin{center}
\subfigure[]{\includegraphics[scale=0.24]{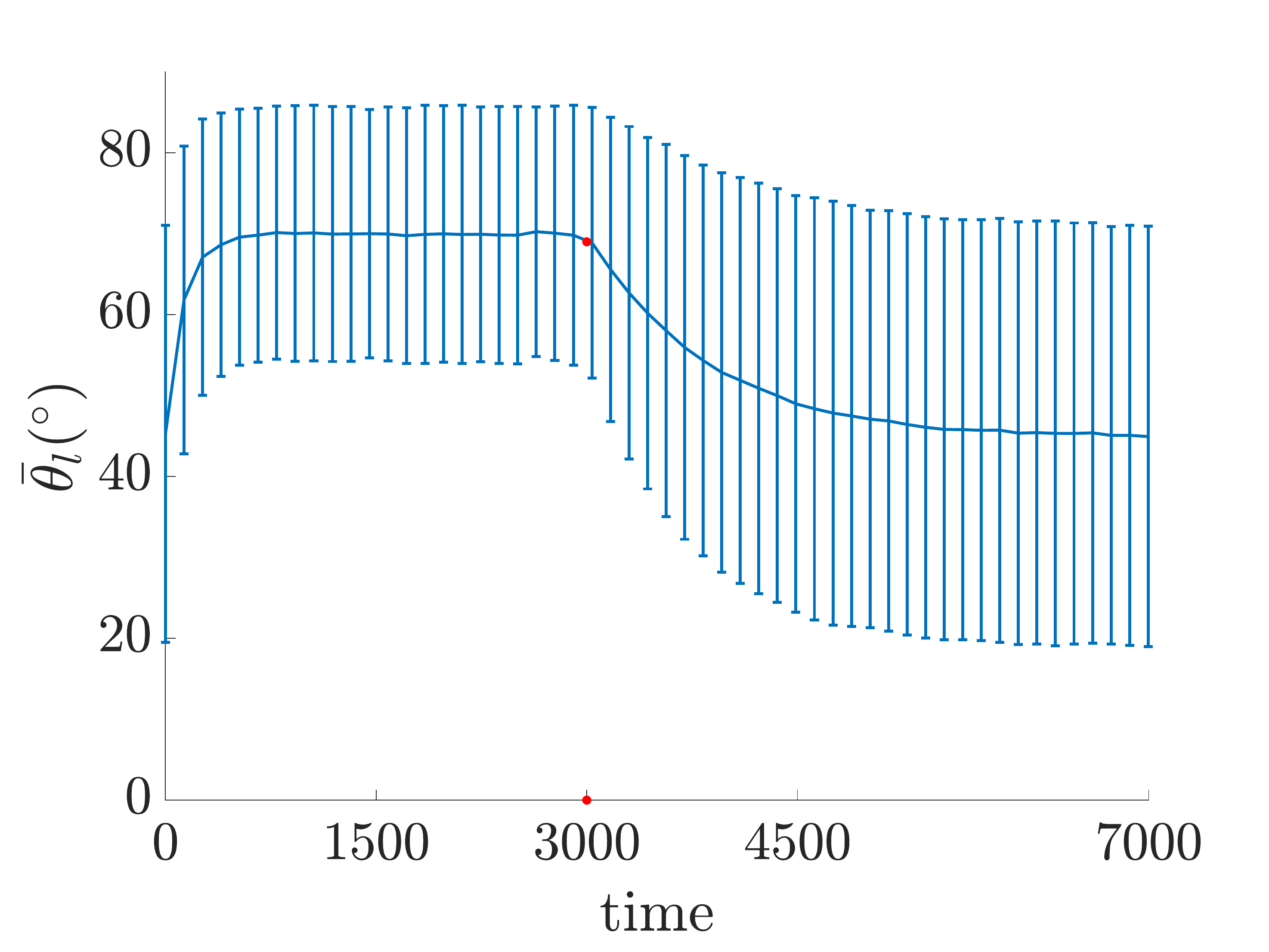}}
\subfigure[]{\includegraphics[scale=0.24]{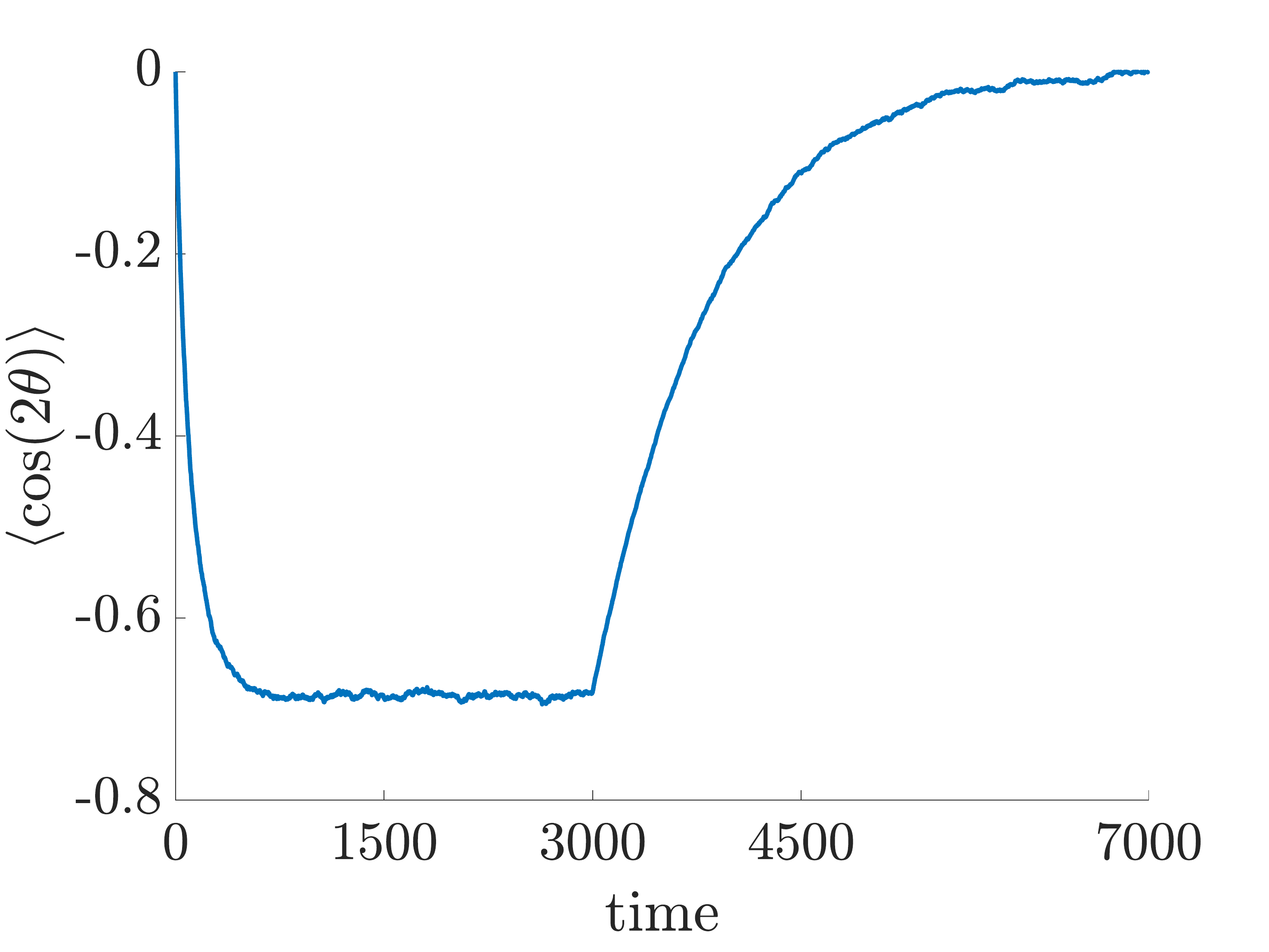}}
\end{center}
\caption{Temporal evolution of the mean of the orientation distribution. In (a) $\omega=1.2$\,Hz and $\varepsilon=10\%$ as reported in \cite{Livne}. In addition, $\lambda_{\theta}=6.6$\,s and  $\sigma=0.7$.
In (b) and (c) $\omega=2$\,Hz and $\varepsilon=8\%$ as reported in \cite{Jungbauer}. In addition, $\lambda_{\theta}=6.6$\,s and  $\sigma=1.6$. After 3000 seconds stretching stops and cells tend to re-orient uniformly over the angle.
Standard deviation of the angle is also given in  (b). In (c) the same mean is reported in terms of its $\cos 2\theta$ for a more direct comparison with \cite{Jungbauer}.  
}
\label{fig_Livne}
\end{figure}

Focusing on the temporal evolution of \eqref{eq:FP_model2} in Fig. \ref{fig_Livne}a we report the results obtained performing a Montecarlo simulation of \eqref{eq:model2} with $N=10^6$ elements, $\gamma=10^{-2}$. In particular, we choose the data of the experimental results reported by Livne et al. \cite{Livne} where $\varepsilon=10\%$, $\lambda_{\theta}=6.6$\,s and $\omega=1.2$\,Hz, corresponding to a high frequency regime,  and we set $\sigma=0.7$ so that the average orientation $\bar{\theta}_l$ of \eqref{eq:stat_mod2}  with $\bar{\sigma}^2=\dfrac{\sigma^2}{2\e^2}$ is the same as reported in \cite{Livne}. The qualitative behaviour corresponds to that reported in \cite{Livne}. In particular we find that the rotation time is $\lambda_{\theta}/\e^2$ as stated in \cite{Livne}.

Finally, we want to replicate the experiment proposed by Jungbauer \cite{Jungbauer} who stop stretching at a certain time and record the recovery phase toward a uniform distribution. To this aim, in Fig. \ref{fig_Livne}b,c the stretch is imposed only for 3000 seconds, while $\e=8\%$ and $r=0.194$. We choose the same re-orientation time as found in \cite{Livne}, i.e. $\lambda_{\theta}=6.6$\,s.
Also in this case the behaviour corresponds to that reported in \cite{Jungbauer}.

\section{Discussion}

In order to describe the dynamics of cell re-orientation under stretch, we proposed a class of Fokker-Planck models, paying particular attention to their link with the microscopic rules.  
In particular, we introduce a local and a non-local rule related to the evaluation of the state of stress experienced by the cell extending its protrusions, and a model of re-orientation as a result of an optimal control activated by the cell. 
The model is able to describe both the evolution and the stationary state of the probability density function over the orientations of the cells, which can be determined explicitly. 
The results compare well with several indipendent experiments \cite{Faust, Hayakawa, Jungbauer, Livne, Mao} showing 
the flexibility of the model.

At present, the microscopic dynamics determining the drift term in the Fokker-Planck equation is defined according to biophysically sound qualitative arguments. But, in the future the close link between the microscopic and the mesoscopic model shown here can be exploited on the one hand to better calibrate the model with respect to experimental data and on the other hand to  describe the microscopic mechanisms starting from measurements on  the behaviour of single cells, whenever these data will be experimentally available.

\section*{Acknowledgements}
This work was partially supported by MIUR (Italian Ministry of Education, Universities and Research) through the PRIN project n. 2017KL4EF3 on ``Mathematics of active materials: From mechanobiology to smart devices'' and through the ``Dipartimento di Eccellenza" 2018–2022 project n. E11G18000350001. We also acknowledge the use of DISMA computational resources.  
NL is a postdoctoral research fellow (“titolare di Assegno di Ricerca”) of Istituto Nazionale di Alta Matematica (INdAM, Italy) also acknowledging support from the National Group of Mathematical Physics (GNFM) grant ``Progetto Giovani 2020".

The authors declare that they have no known competing financial interests or personal relationships that could have appeared to influence the work reported in this paper.

\bibliographystyle{plain}
\bibliography{KineticStressCell}
\end{document}